\documentclass{aa} 
\usepackage{graphicx} 
\usepackage{times}
\usepackage{natbib} 
\bibpunct{(}{)}{;}{a}{}{,} 

\newcommand{\msun}{$\mathrm{{M}_\odot}$}

\newcommand{\mjup}{$\mathrm{{M}_J}$}
\newcommand{\mjups}{$\mathrm{{M}_J}$ }
\newcommand{\mpl}{$M_{\mathrm p}$}

\begin{document}
  \title{RODEO: a new method for planet-disk interaction} 
  \author{Sijme-Jan Paardekooper \and Garrelt Mellema} 
  \author{Sijme-Jan Paardekooper \inst{1} \and Garrelt Mellema \inst{2,1}} 
  \offprints{S. J. Paardekooper\\\email{paardeko@strw.leidenuniv.nl}} 
  \institute{Leiden Observatory, Postbus 9513, NL-2300 RA Leiden, 
             The Netherlands \and
             ASTRON, Postbus 2, NL-7990 AA Dwingeloo, The Netherlands \\
    \email{paardeko@strw.leidenuniv.nl; gmellema@astron.nl}} 

  \date{Draft Version \today} 
  
  \abstract{In this paper we describe a new method for studying the
    hydrodynamical problem of a planet embedded in a gaseous disk.
    We use a finite volume method with an approximate Riemann solver (the
    Roe solver), together with a special way to integrate the source
    terms. This new source term integration scheme sheds new light on
    the Coriolis instability, and we 
    show that our method does not suffer from this instability.The first 
    results on flow structure and gap formation are
    presented, as well as accretion and migration rates. For \mpl $ < 0.1$ 
    \mjups and \mpl $ > 1.0$ \mjups (\mjups = Jupiter's mass) the accretion
    rates do not depend sensitively on numerical parameters, and we find
    that within the disk's lifetime a planet can grow to $3-4$ \mjup. In 
    between these two limits numerics play a major role, leading to 
    differences of more than $50 \%$ for different numerical parameters. 
    Migration rates are not affected by numerics at all as long as the mass
    inside the Roche lobe is not considered. We can reproduce
    the Type I and Type II migration for low-mass and high-mass planets,
    respectively, and the fastest moving planet of $0.1$ \mjups has a 
    migration time of only $2.0~10^4$ yr.
    
     }
  
  \maketitle
  
\section{Introduction}
It is now generally believed that planets are formed out of the same
nebula as their parent star. When this cloud of gas collapses into a
protostar, conservation of angular momentum leads to the formation of
an accretion disk around the star. These disks are indeed observed
around T-Tauri stars \citep{1996Natur.383..139B}, and it is within
these disks that planet formation should take place.

In standard theory, terrestrial planets as well as the rocky cores of
gas giant planets arise slowly from collisions of dust particles.
When a protoplanet reaches a certain critical mass ($\approx 15$ $\rm
M_\oplus$), it can no longer sustain a hydrostatic atmosphere, and
dynamical accretion sets in.  Eventually, this accretion process forms
a gaseous envelope around the core, of a mass comparable to
Jupiter. This scenario is known as the 'core accretion model'. 

Recently, \citet{1997Sci...276.1836B} revisited a model originally
proposed by \citet{1978M&P....18....5C}, the 'core collapse scenario',
in which a giant planet is formed in much the same way as a star
through a gravitational instability in the disk. However, 
\cite{2005ApJ...621L..69R} pointed out several problems with this 
scenario, and it is yet undecided
which of these two scenarios is correct. In both cases we end up
with a newly formed giant planet still embedded in the protoplanetary
disk. It is this interesting stage of planet formation that we focus
on in this paper.

The disk and the planet interact gravitationally with each other. The
planet perturbs the disk through tidal forces, breaking its
axisymmetry and, if the planet is massive enough, opens up a gap in
the disk \citep{1993prpl.conf..749L}. The perturbed disk in its turn
exerts a torque on the planet, leading possibly to orbital migration
\citep{1980ApJ...241..425G}.  Inward migration is generally believed
to be the mechanism responsible for creating 'Hot Jupiters': giant
planets, comparable to Jupiter, with very small semi-major axes ($<
0.1$ AU).

The process of gap formation raises two interesting questions. First:
does the presence of the gap prevent further gas accretion? If it
does, this puts serious constraints on the maximum planet mass that
can be reached.  Secondly: how is the orbital evolution of the planet
affected by the gap? If all planets move inward on much shorter time
scales than the typical lifetime of the disk, how come we
still have Jupiter in our Solar system at $5.2$ AU?

The accretion issue has been investigated in detail numerically by
\citet{1999MNRAS.303..696K} and by \citet{1999ApJ...526.1001L}, and
they find that accretion can continue through the gap, allowing more
massive planets to be formed. Migration has been investigated both
analytically \citep{1980ApJ...241..425G,1986ApJ...309..846L}, recently 
by \citet{2002ApJ...572..566R}, and numerically
by \citet{2000MNRAS.318...18N} and \citet{2000MNRAS.313L..47K}, and it
seems that migration is always inward, allowing for Hot Jupiters but
not for `Cold' Jupiters. Recently \cite{2003ApJ...588..494M} and
\cite{2004ASPC..324...39A} showed that there exists a runaway regime 
in which the orbital radius of the planet evolves on very short timescales, 
with possible outward migration. 

However, the problem is complicated analytically, and hard to do
numerically as well. A nice illustration of this is given by
\citet{1998A&A...338L..37K}, who showed that the use of a corotating
coordinate frame can lead to non-physical evolution in some numerical
codes. It therefore seems appropriate to introduce a numerical method 
which is new in this field of research, and which should handle rotating
coordinate frames and the effects of gravity in a better way. We use a
finite-volume method based on an approximate Riemann solver for
arbitrary coordinate frames, which can is specifically aimed to treat
discontinuities in the flow correctly. Also, the numerical
scheme is conservative, meaning that the method conserves mass,
momentum and angular momentum up to second order.

In this paper, we aim to give a full description of RODEO (ROe solver
for Disk-Embedded Objects), to show some differences in the gas flow in the 
disk and to present results on accretion and migration rates. The focus
will be more on numerical effects than on physical effects, i.e. we do
not consider different disk structures or different magnitudes of 
an anomalous viscosity as was done by \cite{1999ApJ...514..344B} and
\cite{1999MNRAS.303..696K}. Instead, we consider the effect of various 
numerical parameters to assess their importance regarding gap formation,
accretion and migration rates.

We focus on the two-dimensional, 
vertically integrated problem, which is formally only correct for planets 
for which the Roche lobe is larger than the disk scale height. For our
disk, this comes down to \mpl $>$ \mjup. However, because two-dimensional
simulations are far less expensive in terms of computing time than 
three-dimensional simulations this allows us to do a detailed numerical
study and compare our findings to previous two-dimensional results.

We start in Sect. \ref{secPhys} by describing the 
physical model that was used. Next, we focus on the numerical method in 
some detail in Sect. \ref{secNum}. In Sect. \ref{secTest} we show the 
results of some test problems and in Sect.  \ref{secMod} we give the 
initial and the boundary conditions. We give our results on gap formation,
accretion and migration in Sects. \ref{secGap}, \ref{secAcc} and \ref{secMig}.
Section \ref{secSum} is reserved for a short summary and the conclusions.

\section{Physical Model}
\label{secPhys}
\subsection{Basic equations}
Protoplanetary disks are fairly thin, i.e. the ratio of the vertical
thickness $H$ and the radial distance from the center $r$ is smaller
than unity. Typically $H/r = 0.05$. We can therefore vertically integrate
the hydrodynamical equations, and work with vertically averaged state
variables. We will use cylindrical coordinates $(r,\phi)$, with the
central star of $1$ $\mathrm{M_\odot}$ located at $r=0$.

The flow of the gas is determined by the Euler equations, which
express conservation of mass, momentum in all spatial directions, and
energy. These equations can be written in a simple form:
\begin{equation}
\label{eqEuler}
\frac{\partial \vec{W}}{\partial t} + \frac{\partial \vec{F}}{\partial
r}+ \frac{\partial \vec{G}}{\partial \phi} = \vec{S},
\end{equation}
where $\vec{W}$ is called the state, $\vec{F}$ and $\vec{G}$ are the
radial and the polar flux, respectively, and $\vec{S}$ is the source
term. The state consists of the conserved quantities, and
Eq. (\ref{eqEuler}) expresses the fact that any change in time of the
state in a specific volume is due to flow across the boundary of the
volume (the flux terms) or due to a source inside the volume (the
source term). We will omit the energy equation for now, because we
will work with a simple (isothermal) equation of state, which does not
require an energy equation. The method still includes the energy
equation as an option, however, and this way we have another way of
simulating isothermal flow: a run with a very low adiabatic exponent
$\Gamma$, as was done for example in \citet{2003ApJ...589..556N}. We
will present the method only for the isothermal equations; for the
full method including the energy equation we refer to
\citet{1995A&AS..110..587E}.

In cylindrical coordinates, the vectors $\vec{W}$, $\vec{F}$,
$\vec{G}$ and $\vec{S}$ can be written in the following form:
\begin{eqnarray}
\vec{W}=r(\Sigma,~\Sigma v_\mathrm{r},~\Sigma v_\mathrm{\phi}) \\ 
\vec{F}=r(\Sigma v_\mathrm{r},~
\Sigma v_\mathrm{r}^2+p,~ \Sigma v_\mathrm{r} v_\mathrm{\phi})\\ 
\vec{G}=r(\Sigma v_\mathrm{\phi},~ \Sigma
v_\mathrm{r} v_\mathrm{\phi},~ \Sigma v_\mathrm{\phi}^2+p/r^2)
\end{eqnarray}
\begin{eqnarray}
\vec{S}=\left( \begin{array}{c} 0 \\
\Sigma r^2 (v_\phi+\Omega)^2 - \Sigma r
\frac{\partial{\Phi}}{\partial{r}} + p \\ -2 \Sigma v_\mathrm{r}
(\Omega+v_\phi) - \frac{\Sigma}{r}
\frac{\partial{\Phi}}{\partial{\phi}}
 \end{array}\right)
\label{eq2D}
\end{eqnarray}
The symbols used are listed in Table \ref{tabSym}.\\
\begin{table}
\label{tabSym}
\caption{Definition of symbols}
\begin{center}
$
\begin{array}{ll}
\hline r & $radial coordinate$ \\ 
\phi & $azimuthal coordinate$ \\
\Sigma & $surface density$ \\ 
v_\mathrm{r} & $radial velocity$ \\ 
v_\phi & $angular velocity in the corotating frame$ \\ 
p & $vertically integrated pressure$ \\ 
\Phi & $gravitational potential$ \\ 
\Omega & $angular velocity coordinate system$ \\ 
r_\mathrm{p} & $orbital radius of the planet$ \\
P & $orbital period of the planet$ \\
\end{array}
$
\end{center}
\end{table}
Note that in this form there are no derivatives of dependent variables
in the source term, keeping $\vec{S}$ finite even in the presence of
discontinuities. Also, there are no viscous terms present, because we
deal with viscosity in a separate way (see Sect. \ref{secVisc}).

The gravitational potential contains terms due to the central star and
due to the direct and indirect influence of the planet:
\begin{equation}
\label{eqpot}
\Phi(r,\phi)=-\frac{\rm{GM_\odot}}{r} - \frac{{\rm{G}} M_{\rm{p}}}{r_2} +
\frac{{\rm G}M_{\rm{p}}}{r_{\mathrm p}^2}r \cos(\phi-\phi_{\mathrm p})
\end{equation}
Here $r_2=|r-r_\mathrm{p}|$ denotes the distance to the planet. Because this
potential is singular at the position of the planet, and to account
for the threedimensional structure of the disk, we use a smoothed
version of $r_2$:
\begin{equation}
r_2=\sqrt{r^2+r_\mathrm{p}^2-2r~r_\mathrm{p}~\cos (\phi-\phi_\mathrm{p})+
\epsilon^2} \nonumber \\
\end{equation}
The smoothing parameter $\epsilon$ is taken to be a certain fraction
of the Roche lobe of the planet:
\begin{equation}
\epsilon=b~\mathrm{R_R}=b~r_\mathrm{p}~\sqrt[3]{\frac{M_\mathrm{p}}{3\mathrm{M_\odot}}}
\end{equation}
Throughout this paper we have used $b=0.2$. The results do not depend
much on its value, as long as $b < 1$.
The indirect term in the potential (the last term on the right-hand side 
in Eq. \ref{eqpot}) arises due to the fact that a coordinate system centered 
on the central star is not an inertial frame, because the star feels the 
gravitational pull of the planet. A similar (small) term 
for the disk is omitted for simplicity. 

\subsection{Equation of state}
Equation (\ref{eqEuler}) needs to be complemented by an equation of
state. We assume an isothermal equation of state:
\begin{equation}
p=c_\mathrm{s}^2 \Sigma
\end{equation}
This choice is based on the assumption that the gas is able to radiate
away all its excess energy very efficiently. In vertical hydrostatic 
equilibrium, the isothermal sound speed $c_\mathrm{s}$ is given by:
\begin{equation}
\label{eqsnd}
c_\mathrm{s} = \frac{H}{r} \sqrt{\frac{\mathrm{G~M_\odot}}{r}},
\end{equation}

\subsection{Viscosity}
\label{secVisc}
The nature of the viscosity in accretion disks was unknown for a long time.
Currently, the Magneto-Rotational Instability (MRI) is the best candidate for
providing a turbulent viscosity \citep{1990BAAS...22.1209B}. In regions where
the ionisation fraction in the disk is high enough to sustain the MRI one 
would need to do Magneto-Hydrodynamics (MHD) to study planet-disk interaction, 
as was done in \cite{2003MNRAS.339..993N}. MHD simulations are very expensive 
however, and it is not yet clear if the MRI operates throughout the disk. 
In so-called dead-zones \citep{1996ApJ...457..355G} there may be no turbulent
viscosity at all. However, in order to compare our results to previous
studies we include an anomalous turbulent viscosity in our models.

Viscosity comes in as two extra source terms for the momentum, one in
the radial and one in the azimuthal direction. We deal with these
source terms separately in the numerical method (see
Sect. \ref{secNum}).

The form of these source terms can be found for example in
\citet{2002A&A...385..647D}:
\begin{eqnarray}
f_\mathrm{r}&=&\frac{1}{r} \frac{\partial(r \mathcal{S}_\mathrm{rr})}
{\partial r} +
\frac{1}{r} \frac{\partial(\mathcal{S}_\mathrm{r\phi})}{\partial \phi} -
\frac{\mathcal{S}_{\phi \phi}}{r} \\ f_\phi&=&\frac{1}{r}
\frac{\partial(r^2 \mathcal{S}_\mathrm{r\phi})}{\partial r} +
\frac{\partial(\mathcal{S}_{\phi \phi})}{\partial \phi}
\end{eqnarray}
where the components of the viscous stress tensor $\mathcal{S}$ are:
\begin{eqnarray}
\mathcal{S}_\mathrm{rr}&=&2\nu\Sigma \left(\frac{\partial v_\mathrm{r}}
{\partial r} -
\frac{1}{3} \nabla \cdot \vec{v} \right) \\ \mathcal{S}_{\phi
\phi}&=&2\nu\Sigma \left(\frac{\partial v_\phi}{\partial \phi} +
\frac{v_\mathrm{r}}{r} - \frac{1}{3} \nabla \cdot \vec{v} \right) \\
\mathcal{S}_\mathrm{r \phi}&=&\nu\Sigma \left(\frac{1}{r}\frac{\partial
v_\mathrm{r}}{\partial \phi} + r \frac{\partial v_\phi}{\partial r} \right) \\
\nabla \cdot \vec{v} &=& \frac{1}{r}\frac{\partial(r v_\mathrm{r})}{\partial r}
+ \frac{\partial v_\phi}{\partial \phi}
\end{eqnarray}   
We can either use an $\alpha$-prescription for the viscosity parameter
$\nu$ \citep{1973A&A....24..337S}:
\begin{equation}
\label{eqalpha}
\nu = \alpha c_\mathrm{s} H
\end{equation}
or we can take $\nu$ constant throughout the disk. Note that for the
$\alpha$-disk with constant aspect ratio $H/r$, $\nu \propto
\sqrt{r}$, leading to enhanced viscosity in the outer disk.
 
\section{Numerical Method}
\label{secNum}
We can integrate Eq. (\ref{eqEuler}) over the finite volume of a grid
cell to obtain:
\begin{eqnarray}
\frac{1}{\Delta t \Delta r \Delta \phi}~( \int dr \int d\phi
  ~(\vec{W}^{n+1}-\vec{W}^n) +\nonumber \\ \int dt \int d\phi
  ~(\vec{F}_{i+1/2,j}-\vec{F}_{i-1/2,j}) +\nonumber \\ \int dt \int dr
  ~(\vec{G}_{i,j+1/2}-\vec{G}_{i,j-1/2}) - \nonumber \\ \int dt \int
  dr \int d\phi~ \vec{S}) = 0
\end{eqnarray}
Here $A^{n}_{i,j}$ means the physical quantity $A$, evaluated 
at time index $n$, at coordinates $(i,j)$. The volume term 
($r^2 \sin \theta$ for spherical coordinates, $r$ for cylindrical coordinates) 
of the grid cell is already present in $\vec{W}$, $\vec{F}$ and $\vec{S}$. A
second order accurate integration scheme for this equation is:
\begin{eqnarray}
\frac{1}{\Delta
t}(\langle\vec{W}\rangle^{n+1}-\langle\vec{W}\rangle^{n}) +\nonumber
\\ \frac{1}{\Delta
r}~(\vec{F}^{n+1/2}_{i+1/2,j}-\vec{F}^{n+1/2}_{i-1/2,j}) +\nonumber \\
\frac{1}{\Delta
\phi}~(\vec{G}^{n+1/2}_{i,j+1/2}-\vec{G}^{n+1/2}_{i,j-1/2}) -\nonumber
\\ \vec{S}^{n+1/2}_{i,j}=0
\label{eqintscheme}
\end{eqnarray}
where $\langle \rangle$ denotes arithmetic mean. A numerical scheme
like this is called \emph{conservative} because the conserved
quantities are indeed conserved by the numerical method: what goes
into one cell must come out of another.

We will use Eq. (\ref{eqintscheme}) to find an update for the
state. What is left to be done is to find expressions for the
interface fluxes (Sect. \ref{secRoe}) and to account for the source
terms (Sect. \ref{secStat}).

\subsection{Roe solver}
\label{secRoe}
First, we use the technique of dimensional splitting to obtain the two
one-dimensional equations
\begin{eqnarray}
\label{eq1D}
\frac{\partial{\vec{W}}}{\partial{t}}+\frac{\partial{\vec{F}}}{\partial{r}}=
\vec{X}
\nonumber\\
\frac{\partial{\vec{W}}}{\partial{t}}+\frac{\partial{\vec{G}}}{\partial{\phi}}=
\vec{Y}
\label{eqsplit}
\end{eqnarray}
The order in which these equations are solved is alternated to avoid
systematic numerical effects \citep{1968...............}. We have the
freedom of choosing the splitting of the source term any suitable
way. We discuss a special way in Sect.~\ref{secStat}.

From now on, we focus on the radial direction; the azimuthal
integration is done in a similar way.  We split Eq.~(\ref{eq1D}) once
more to obtain an equation without source terms, and solve this
equation with a method originally proposed by
\citet{1981...............}, and extended by
\citet{1995A&AS..110..587E} to a general relativistic method. The
non-relativistic limit of this method yields a solver for the Euler
equations in arbitrary coordinates.

\subsubsection{Characteristics}
Given the state (or the flux) immediately left and right of an
interface of two grid cells, the Roe solver computes the resulting
interface flux by solving:
\begin{equation}
\label{eqadv}
\frac{\partial\vec{W}}{\partial{t}}+\mathcal{A}(\vec{W})\frac{\partial
\vec{W}}{\partial r}=0
\end{equation}
where $\mathcal{A}(\vec{W})= \partial \vec{F} /\partial \vec{W}$ is
the Jacobian matrix. Roe's central idea is to approximate
$\mathcal{A}(\vec{W})$ by a constant matrix $\tilde
\mathcal{A}$. Since the Euler equations are hyperbolic, this matrix
$\tilde \mathcal{A}$ has eigenvectors $\vec{e}_{k}$ and corresponding
eigenvalues $\lambda_{k}$. These can be used to diagonalize $\tilde
\mathcal{A}$:
\begin{equation}
\mathcal{Q}^{-1}~\tilde \mathcal{A}~ \mathcal{Q} = \mathcal{D}
\end{equation} 
where $\mathcal{Q}$ is the matrix with right eigenvectors of $\tilde
\mathcal{A}$.  Now we can cast Eq. (\ref{eqadv}) into characteristic
form:
\begin{equation}
\label{eqcharac}
\frac{\partial \vec{C}}{\partial t}+ \mathcal{D} \frac{\partial
\vec{C}}{\partial r}=0
\end{equation}
where $\vec{C}$ is the vector of characteristic variables, defined by
\begin{equation}
\label{eqcharac2}
d\vec{C}=\mathcal{Q}^{-1}~d\vec{W}
\end{equation}
and $\mathcal{D}$ is the diagonal matrix with eigenvalues of $\tilde
\mathcal{A}$. From (\ref{eqcharac}) it is easy to see that
$d\vec{C}=0$ along the path $\vec{r}=\mathcal{D} \vec{t}$. These paths
are called characteristics, and are essential in the study of
hyperbolic differential equations.  Discontinuities (shocks) travel
along characteristics, and the domains of influence and dependence are
bounded by the characteristics.

We can integrate Eq. (\ref{eqcharac}) to find a relation between the
state and the characteristic variables:
\begin{equation}
\label{eqproj}
\vec{W}=\mathcal{Q}\vec{C}=\sum_{k} a_{k} \vec{e}_{k}
\end{equation}
where $a_{k}$ is the k-th characteristic variable. As the $a_{k}$ 
are used to project the state on the eigenvectors of $\tilde \mathcal{A}$, 
they are called \emph{projection coefficients}.

We can use the projection coefficients for calculating the interface
flux, because we know that they are constant along characteristics. If
we can figure out from which point in space the characteristics that
cross the interface originate at the current time, we can just
calculate the projection coefficients at that point in space and use
Eq. (\ref{eqproj}) to find the state at the interface.

The interface flux is related to the interface state by:
\begin{equation}
\vec{F}=\tilde \mathcal{A} \vec{W}=\tilde \mathcal{A} \mathcal{Q}
\vec{C} = \mathcal{Q} \mathcal{D} \vec{C} = 
\sum_{k} b_{k} \vec{e}_{k}
\end{equation}
where $b_{k}=\lambda_{k} a_{k}$. We present the exact 
expressions for the eigenvalues, eigenvectors and projection coefficients 
in Appendix \ref{appA}.

We can project the flux difference at the interface we are considering
on the eigenvectors of $\tilde \mathcal{A}$:
\begin{equation}
\vec{F}_{R}-\vec{F}_{L}=\sum b_{k} \vec{e}_{k}
\end{equation}
where $\vec{F}_{R}$ and $\vec{F}_{L}$ are the fluxes immediately
right and left of the interface, respectively. Then the first order 
interface flux is approximated by:
\begin{equation}
\label{eqfirstorder}
\vec{F}^{n+1/2}_{i+1/2} =
\frac{1}{2}(\vec{F}_{L}+\vec{F}_{R})-
\frac{1}{2}\sum \sigma_{k} b_{k}
\vec{e}_{k}
\end{equation}
where $\sigma_{k}=\rm{sign}(\lambda_k)$.

\subsubsection{Flux limiter}
\label{secfluxlim}
The first order expression for the flux, Eq. (\ref{eqfirstorder}),
assumes a jump in the projection coefficients. That is: depending on
the sign of the eigenvalue $\lambda_{k}$, we use the $b_{k}$ 
corresponding to either the left or the right state. This procedure is 
correct if there is a shock right at that interface, so the state makes a jump
there. In regions of smooth flow however, linear interpolation between
the different $b_{k}$ seems to be a better approach. To switch between
the two kinds of interpolation a flux limiter $\psi$ is used. We
follow the suggestion by Roe (1985), in which one uses the ratio of
the state difference at the interface $a_0=[a_{k}]_{i+1/2}$ to the
upwind state difference:
\[
a_u=\left\{
\begin{array}{ll}
  (a_{k})_{i-1/2} & \mbox{if $\lambda_{k} \geq 0$};\\ 
(a_{k})_{i+3/2} &
  \mbox{if $\lambda_{k} < 0$}
\end{array} \right.
\]
Let
\begin{small}
\begin{eqnarray}
\psi (a_0,a_u)=
\max (0,\min (p a_u, \max (a_0,\min (a_u,p a_0))))+
               \nonumber \\ \min (0,\max (p a_u, \min (a_0,\max
               (a_u,p a_0))))
\label{eqfluxlim}
\end{eqnarray}
\end{small}
This form of the flux limiter allows us to tune the way the method
deals with sharp discontinuities through the parameter $p$. When 
$p = 2$, for example, the flux limiter is called
'Superbee'. This limiter tends to create very sharp shocks, but tends
to steepen shallow gradients, leading to numerical problems
such as under- and overshoots. For $p = 1$ the limiter is
called 'minmod', which is more diffusive. To get the best of both
worlds (sharp shocks but no under- or overshoots), we set $p = 1.5$ in 
all simulations.

With this flux limiter, the second order interface flux becomes:
\begin{equation}
\label{eqsecorder}
\vec{F}^{n+1/2}_{i+1/2} =
\frac{1}{2}(\vec{F}_{L}+\vec{F}_{R})-
\frac{1}{2}\sum (\sigma_{k} a_{k}
-(\sigma_{k}-\nu_{k})\psi_{k}) \lambda_{k} \vec{e}_{k}
\end{equation}
where $\nu_{k} = \lambda_{k} \Delta t/\Delta r$, the coefficient 
needed for linear interpolation. When $\psi_{k}=0$ (shock),
Eq. (\ref{eqfirstorder}) is retrieved, while for $\psi_{k}=1$ (smooth
flow) the $\sigma_{k}$ in Eq. (\ref{eqfirstorder}) is replaced by
$\nu_{k}$ and we interpolate linearly between the projection
coefficients.

\subsection{Stationary Extrapolation}
\label{secStat}
The usual way for dealing with source terms is to split Eq. \ref{eqsplit}
once more to end up with an ordinary differential equation for the state: 
\begin{equation}
\label{eqHom}
\frac{\partial \vec{W}}{\partial t} = \vec X,\quad \frac{\partial
\vec{F}}{\partial r} = 0
\end{equation}
thereby assuming that the flux is constant in space. The ordinary differential
equation can be solved with any second-order integration scheme to yield a
second-order update for the state. 

However, it is also possible to take the opposite approach:
\begin{equation}
\label{eqstat}
\frac{\partial \vec{W}}{\partial t} = 0, \quad \frac{\partial
\vec{F}}{\partial r} = \vec X
\end{equation}
This is a method we call \emph{stationary extrapolation}, because it
assumes a stationary state within one grid cell. This equation is more
difficult to solve, but it has certain advantages:
\begin{itemize}
\item 
The Roe solver solves a Riemann problem at the interface, a
configuration with two \emph{stationary} states separated by a
discontinuity. Using actual stationary states on either side of the
interface ensures that the Roe solver deals with a genuine Riemann
problem.
\item
Physical stationary states are recognized. When the actual states
\emph{are} stationary, the Roe solver will produce no (unwanted)
numerical evolution of these states. This property is related to the 
numerical instability found by \cite{1998A&A...338L..37K} concerning the
Coriolis force. We will see below that a scheme using stationary extrapolation 
does not suffer from this instability.
\end{itemize}
If we split the source terms appropriately:
\begin{equation}
\vec{X}=\left(\begin{array}{c} 0 \\ \Sigma r^2(v_\phi+\Omega)^2 -
\Sigma r \frac{\partial \Phi}{\partial r} + p \\ -2 \Sigma v_{\rm{r}}
(\Omega+v_\phi)
\end{array} \right)
\end{equation}
\begin{equation}
\vec{Y}=\left(\begin{array}{c} 0 \\ 0 \\ -\frac{\Sigma}{r}
\frac{\partial \Phi}{\partial \phi}
\end{array} \right)
\end{equation}
the integrations can be done analytically \citep{1995A&AS..110..587E}.
For isothermal, stationary flow in the radial direction Eq \ref{eqstat}
can be rewritten:
\begin{equation}
\frac{\partial}{\partial r}\left(\begin{array}{c} r \Sigma v_{\rm r} \\ 
\frac{1}{2}(v_{\rm r}^2 + r^2 v_\phi^2)+\Phi+c^2\log(\Sigma) \\
r^2(v_\phi+\Omega)
\end{array} \right)=0
\end{equation}
That is: the mass flux, the Bernouilli constant and the specific angular 
momentum are invariant. From these invariants the state at a cell interface 
can be computed from the state at the cell center. However, this procedure 
is computationally expensive, therefore it is feasible to adopt a first order 
approximation to calculate the interface fluxes:
\begin{eqnarray}
\vec{F}_{i-1/2,R} = \vec{F}_{i} - 
\frac{1}{2} \Delta r \vec{X}_{i} \nonumber \\ 
\vec{F}_{i-1/2,L} = \vec{F}_{i-1} + \frac{1}{2} \Delta r
\vec{X}_{i-1}
\label{eqapproxstat}
\end{eqnarray}
This procedure can conserve stationary states up to second order 
\citep{1991A&A...252..718M}. However, \cite{1998A&A...338L..37K} showed that
the angular momentum in disk simulations deserves special attention. We 
illustrate this with a simple example. Consider the radial transport of
angular momentum:
\begin{equation}
\frac{\partial}{\partial t}(r \Sigma v_\phi)+
\frac{\partial}{\partial r}(r \Sigma v_r v_\phi)=
-2 \Sigma v_r (v_\phi+\Omega)
\end{equation}
For stationary radial flow the mass flux and the specific angular momentum
are constant: $r \Sigma v_r=D$ and $r^2(v_\phi+\Omega)=L$. For simplicity we 
assume here that $D$ is constant for the whole computational domain. 
Consider the interface between cells $i$ and $i-1$. Stationary extrapolation
from the cell centers to the interface $i-1/2$ leads to the fluxes
\begin{equation}
F_L=D(\frac{L_{i-1}}{r_{i-1/2}^2}-\Omega)
\end{equation}
\begin{equation}
F_R=D(\frac{L_{i}}{r_{i-1/2}^2}-\Omega)
\end{equation}
With $v_r>0$, the first order interface flux produced by the Roe solver is
$F_{i-1/2}=F_L$. Therefore 
\begin{equation}
L_{i-1/2}=r_{i-1/2}^2(F_L/D+\Omega)=L_{i-1}
\end{equation}
Stationary extrapolation from the interface back to the cell center $i$ gives 
the final flux coming from the left for cell $i$:
\begin{equation}
F_{L,i}=D(\frac{L_{i-1}}{r_{i}^2}-\Omega)
\end{equation}
From the same analysis for interface $i+1/2$ we find the flux going to the
right for cell $i$:
\begin{equation}
F_{R,i}=D(\frac{L_{i}}{r_{i}^2}-\Omega)
\end{equation}
The first order update for the state is:
\begin{equation}
\Delta(r\Sigma v_\phi)=\frac{\Delta t}{\Delta r}(F_{L,i}-F_{R,i})=
\frac{\Delta t}{\Delta r}\frac{D}{r_{i}^2}(L_{i-1}-L_i)
\end{equation}
The change in angular momentum $J=r^2\Sigma(v_\phi+\Omega)$ due to this 
update is given by:
\begin{equation}
\Delta J = \frac{\Delta t}{\Delta r}\frac{D}{r_{i}}(L_{i-1}-L_i)
\end{equation}
It is easy to see that the conservative formulation of 
\cite{1998A&A...338L..37K}:
\begin{equation}
\frac{\partial}{\partial t}(r \Sigma L)+
\frac{\partial}{\partial r}(r \Sigma v_r L)=0
\end{equation}
leads to the same change in angular momentum. This shows that this new method
does not suffer from the numerical instability due to the explicit 
treatment of the Coriolis force that was noted by \cite{1998A&A...338L..37K}.
Note however that this does not hold for the approximate extrapolation of
Eq. \ref{eqapproxstat}. Therefore we always use the exact extrapolation for
the specific angular momentum, while keeping the method of approximate 
extrapolation to deal with the other source terms. This way we do not need
to do the computationally intensive full stationary extrapolation, while
keeping angular momentum nicely conserved.

Stationary extrapolation therefore provides a different point of view on the 
Coriolis instability: the failure of numerical hydrodynamic schemes to
properly conserve angular momentum can be traced back to the failure to 
recognize stationary states. Rewriting the angular momentum equation as
done by \cite{1998A&A...338L..37K} is a way to solve this problem, but it
only deals with the Coriolis source term while similar problems may exist
due to the other source terms as well. Therefore we feel that it is a good
idea to integrate \emph{all} source terms using stationary extrapolation.

One disadvantage of stationary extrapolation is that it is not known
a-priori if the assumption of stationary flow is valid. This is
especially important when the source terms are large, in our case very
close to the planet. When dealing with more massive planets ($>0.5$ \mjup) 
the assumption of stationary flow along the coordinate axes
breaks down, leading to numerical instabilities. A physical
explanation is that the gas in this case will try to orbit the planet,
perhaps forming a Keplerian disk, rather than to orbit the star. The
flow is still semi-stationary, but only in a cylindrical coordinate frame
centered on the planet, and this flow is at some points even
orthogonal to the stationary flow in the frame of the star. Therefore,
it seems appropriate to treat these larger source terms as external
(see section \ref{extsource}). This transition from stationary to
non-stationary extrapolation is applied smoothly at a typical distance 
$R_{\rm R}$ from the planet with a $\sin^2$ ramp.

\subsubsection{External source terms}
\label{extsource}
The stationary extrapolation deals with the geometrical source terms
(including gravity, which is merely a geometrical effect in general
relativity). Any other source term which we will call $\vec Z$ is
better integrated using Eq. (\ref{eqHom}):
\begin{equation}
\label{eqHomnum}
\textbf{W}^{n+1} = \textbf{W}^n + \Delta t \textbf{Z}
\end{equation}
In our case, $\vec Z$ consists of the viscous source terms. The
derivatives in these terms are calculated using a finite-difference
method, and the resulting source is fed into Eq. (\ref{eqHomnum}).

\subsection{Adaptive Mesh Refinement}
Resolution is always an issue in numerical hydrodynamics, and 
a compromise has to be found between resolution and computational cost.
Adaptive Mesh Refinement (AMR) is a very economic way of refining your
grid where the highest resolution is needed, whereas keeping large
parts of the grid at low resolution. 

We have implemented the PARAMESH algorithm \citep{2000CoPhC.126..330M}
to be able to resolve the region near the planet accurately. Usually 
the refinement criterium is taken to be the second derivative of the 
density, but we can also predefine the region that has to be refined.
When running in this mode, and switching off additional refining and 
derefining, the algorithm works like the nested grid technique of
\cite{2002A&A...385..647D}. However, since our grid can be fully adaptive
it is suited to let the planet migrate through the disk while keeping
a high resulution near the planet. This will be the subject of a future 
paper.

We use linear interpolation to communicate boundary cells between the 
different levels of refinement. The implementation of the monotonised 
harmonic mean \citep{1977JCoPh..23..276V} used by \cite{2002A&A...385..647D}
made no significant difference. For every boundary between different levels
we reset the flux into the lower level in order to conserve mass and momentum
across the boundary.

\subsection{Accretion}
We follow \citet{2002A&A...385..647D} in modeling the accretion by
taking away some fraction of the density within a distance of 
$r_\mathrm{acc}$ of the planet. Basically,
the density is reduced with a factor
\begin{equation}
\label{eqaccproc}
1-f~ \Delta t,
\end{equation}
where $\Delta t$ is the magnitude of the time step. The two parameters
that determine the accretion rate are $r_\mathrm{acc}$ and $f$, where
$f$ is taken to be twice its value in the inner half of $r_\mathrm{acc}$. 
The mass taken out is monitored, assuming it has been accreted
onto the planet, but without changing the dynamical mass of the
planet.

Equation \ref{eqaccproc} can be seen as a first order approximation to 
the solution of the differential equation
\begin{equation}
\frac{d \rho}{dt}=-\frac{\rho}{\tau_{\rm acc}}
\end{equation}
with the solution
\begin{equation}
\rho(t+\Delta t)=\rho(t) e^{-\Delta t/\tau_{\rm acc}}\approx 
\rho(t)\left(1-\Delta t/\tau_{\rm{acc}}\right)
\end{equation}  
From this equation we see that the typical time scale $\tau_{\rm{acc}}$ 
for emptying the accretion region is $f^{-1}$.

\begin{figure}
\resizebox{\hsize}{!}{\includegraphics[bb=0 0 465 370]{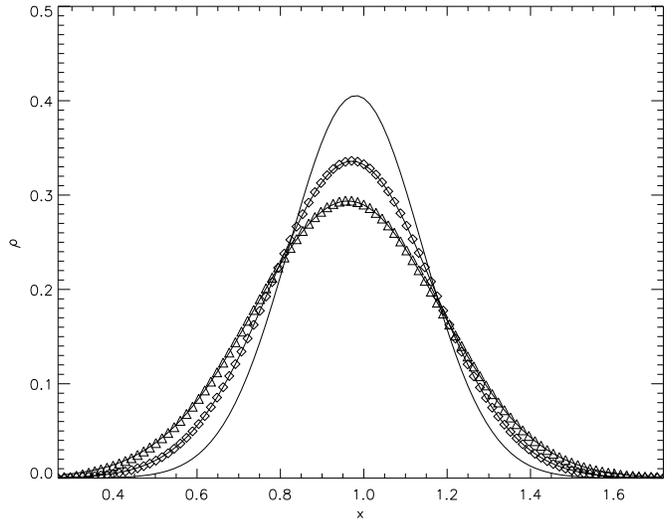}}
\caption{The evolution of the axisymmetric gas ring, initially located
at x=1, shown at $\tau=0.05$, and after 50 and 100 orbits. For the
last two the numerical results are also plotted (diamonds and
triangles, respectively.}
\label{fig1}
\end{figure}

\section{Test Problems}
\label{secTest}
We tested the RODEO method against several hydrodynamical test problems.
First of all, the standard Sod shocktube \citep{1978JCoPh..27....1S},
in one and in two dimensions.  For the two-dimensional problem, we
have placed the initial discontinuity diagonal along the grid, making
it a genuine two-dimensional problem. The analytic solution was nicely
recovered, except at the boundaries where some reflections were seen.

A more complicated test is the windtunnel with step
\citep{1984JCoPh..54..115W}, which was previously used as a
testproblem by \citet{1991A&A...252..718M} using the same method.  The
results agreed very well. Note that both tests do use an energy
equation.

\subsection{Viscous ring spreading}
To test the implementation of the viscous terms, we set up an
axisymmetric ring of gas in Keplerian rotation about a central
object. If we neglect pressure forces, the evolution of the surface
density, which initially is a delta function, is given by
\citet{1981ARA&A..19..137P}:
\begin{equation}
\label{eqvisc}
\Sigma(r,t)=\frac{1}{\pi r_0^2 \tau x^{1/4}}
e^{-\frac{1+x^2}{\tau}}~I_{1/4} \left(\frac{2x}{\tau}\right)
\end{equation}
where x denotes the normalized radial coordinate $r/r_0$, with the
ring initially located at $r=r_0$, and $\tau$ denotes the
dimensionless time ($\tau=t/t_v$) in units of the viscous spreading
time $t_v=r_0^2/12\nu$. $I_{1/4}$ denotes the Bessel function of the
first kind of fractional order $1/4$.

As the analytical solution neglects any forces due to a pressure
gradient, we have to make sure that we set the temperature to a very
low value, or else pressure waves will dominate the pattern.  We have
used the same resolution as for simulations with the planet, and with
a viscosity parameter $\alpha=0.004$ at $x=1$, comparable to the
value used in the disk-planet simulations.  We have placed the ring in
the middle of the computational domain of $x \in [0.27,1,73]$,
with $\tau=0.05$ initially because we cannot model a delta function
numerically.

We have also performed a test run of this problem using $\alpha=0$ to
check for numerical viscosity, and it turned out that this (unwanted)
numerical diffusion was very much lower than any physical viscosity.
at a low resolution of 128x384. Note that this also shows that we can
maintain a stable disk in Keplerian rotation. This is a nice result,
showing that our conservative scheme conserves angular momentum very
well.

Figure \ref{fig1} shows the results for $\alpha=0.004$.  It is
clear that the numerical solution agrees very well with
Eq. (\ref{eqvisc}), indicating that the implementation of the
viscosity works. Only at the boundaries there are minor deviations
from the analytical solution.

\begin{figure*}
\includegraphics[bb=60 275 533 566,width=17cm]{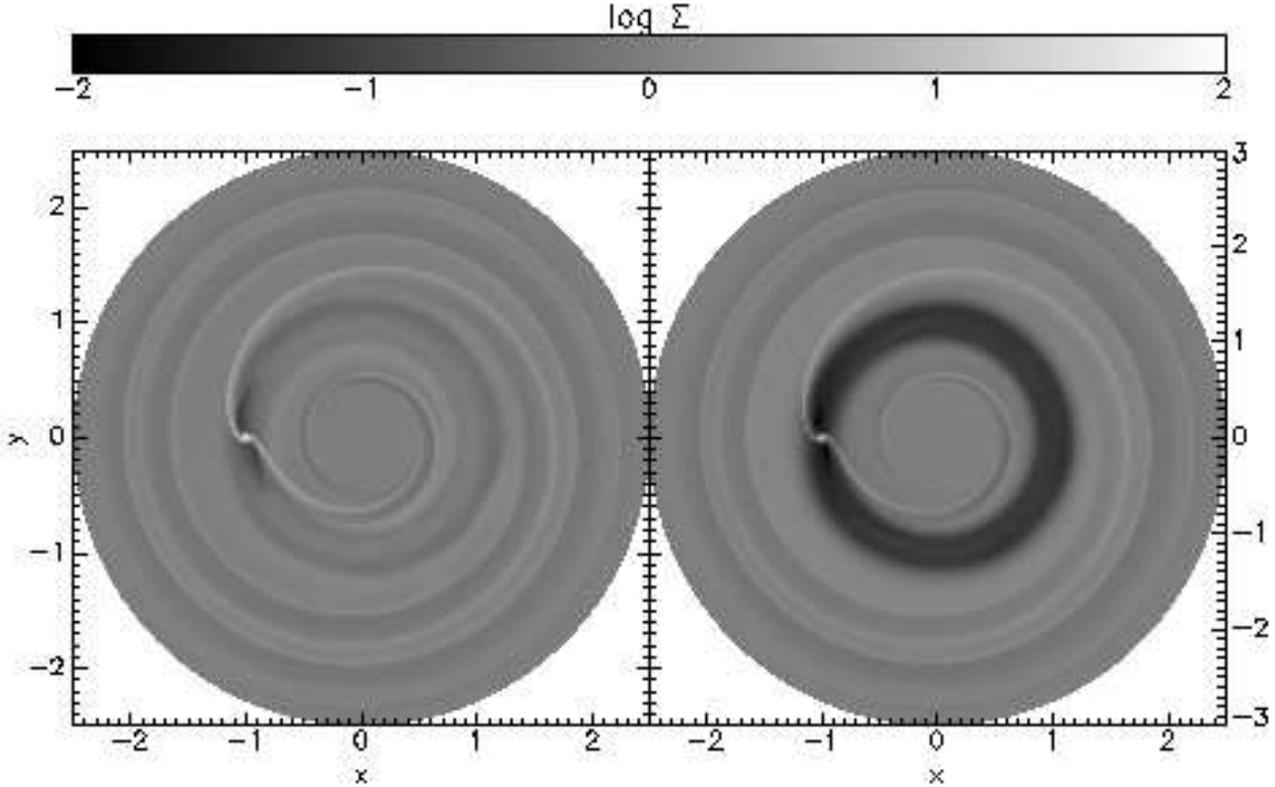}
\caption{Greyscale plot in $(x,y)$ after $10$ orbits (left panel) and
after $200$ orbits (right panel) of a 1 \mjups planet.}
\label{fig2}
\end{figure*}

\section{Model design}
\label{secMod}
Throughout we use non-dimensional units. The unit of distance is the
orbital radius of the planet, and the unit of time is the inverse 
orbital frequency. Note that this differs by a factor $2 \pi$ from 
the orbital period of the planet.

\subsection{Initial Conditions}
The base resolution in all our simulations is 
$(n_\mathrm{r},n_\mathrm{\phi})=(256,768)$, which leads to square
cells near the position of the planet. 
We set $h=H/r=0.05$, which determines the temperature through
Eq. \ref{eqsnd}. We take a constant initial surface density. Because
we do not consider the self-gravity of the disk the density can be given in
any desirable unit, and we normalize it to $1$ at the planet's position. 
We set the radial velocity to zero, although another possibility is to take an 
exact $\alpha$-disk initially. The initial angular velocity is set to the 
value for a Keplerian disk, with a small correction for the pressure force.

We take a constant kinematic viscosity $\nu=10^{-5}$, corresponding to 
$\alpha=0.004$ at the location of the planet. We put in a planet at location 
$(r,\phi)=(1, \pi)$. This planet can be allowed to accrete matter 
(without changing its dynamical mass).

\subsection{Boundary conditions}
The inner and the outer boundary are located at $0.4$ and $2.5$,
respectively. Imposing proper boundary conditions is not trivial,
because waves are continuously hitting both edges of the grid,
changing the sign of the radial velocity. A standard outflow boundary
assumes that the velocity is always directed outward, and in
combination with outgoing waves this leads to instabilities near the
inner radius of the disk. A reflecting boundary, on the other hand,
will lead to reflected waves traveling into the computational domain,
which interact with the outgoing waves, destroying the flow structure.

Therefore we decided to take a \emph{non-reflecting} boundary,
following \citet{1996MNRAS.282.1107G}.  This boundary treatment is
based on characteristics, and as a result the implementation is
particularly simple in our numerical scheme. The idea is to impose the
boundary conditions on the characteristic variables, rather than on
density, momentum or energy directly. \citet{1996MNRAS.282.1107G}
showed that this leads to a non-reflective boundary.

A ghost cell has to be updated according to incoming characteristics:
when a characteristic leaves the last regular cell into the ghost
cell, the corresponding characteristic variable of the ghost cell is
updated accordingly. When a characteristic enters the ghost cell from
outside the numerical domain, the characteristic variable is updated
using the unperturbed (Keplerian) value. Using the Roe solver, this
comes down to having \emph{two} ghost cells: the outer one is never
updated, so density and velocities remain at the initial values, and 
the inner one serves as the actual ghost cell.  The
latter is automatically updated by the Roe solver using characteristic
variables, thereby following the suggestion by
\citet{1996MNRAS.282.1107G}.

It turns out that these boundary conditions work especially well with 
stationary extrapolation, because then the hydrodynamics part as well as
the boundary prescription are dealing with small fluctuations over a
stationary background state. When we switch to the ordinary integration 
of source terms of Eq. \ref{eqHom} this is no longer true. In this case 
small numerical artefacts can be seen near the boundaries. 

Therefore, as an addition we can add a wave-damping mechanism 
\citep{comparison} that operates in the regions $0.4 < r < 0.5$ and 
$2.1 < r < 2.5$. In these regions the state is relaxed to the initial 
(stationary) state on a time scale varying from infinity at $r=0.5$ and 
$r=2.1$ to 1 orbital period at the location of the boundary. Using this 
prescription, outgoing waves are damped gently as they approach the boundary 
of the computational domain.  

\begin{figure*}
\includegraphics[bb=35 10 515 245,width=17cm]{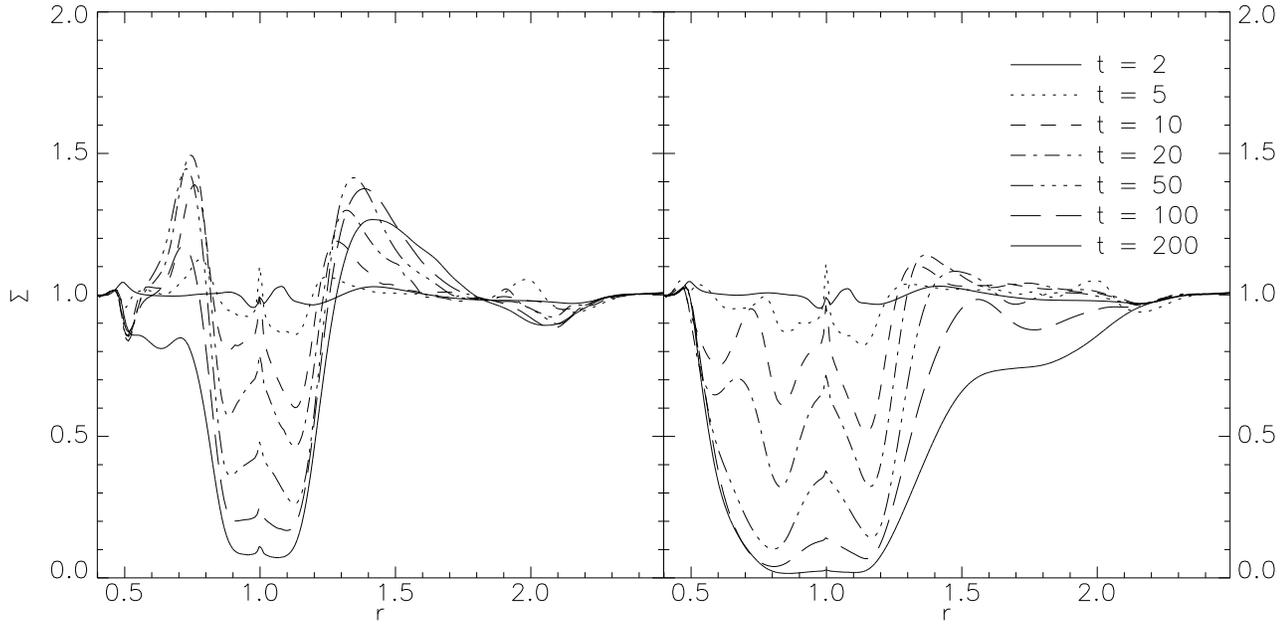}
\caption{Gap formation for a 1 \mjups planet. Left panel: azimuthally 
averaged surface density for the case of exact stationary extrapolation. 
Right panel: azimuthally averaged surface density for the case of 
approximate extrapolation. The indicated times are in planetary orbits.}
\label{fig3}
\end{figure*}

\section{Gap Formation}
\label{secGap}
Only massive planets are able to open gaps in gas disks. There are
two criteria that have to be fullfilled: firstly, the torques arising due 
to the presence of the planet must be able to overcome the viscous torques,
leading to a minimum mass of \citep{1999ApJ...514..344B}:
\begin{equation}
M_\mathrm{min,\nu}=\frac{40 \nu}{\Omega_{\mathrm K} r_{\mathrm p}^2} 
\mathrm{M_\odot}
\end{equation}
Secondly, \cite{1993prpl.conf..749L} suggested that at a distance of 
$\mathrm{R_R}$ from the planet the planet's gravity should be more important
than pressure, which leads to another minimum mass for gap opening:
\begin{equation}
M_\mathrm{min,h}=3~h^3~\mathrm{M_\odot}
\end{equation}
For our disk, with $\nu=10^{-5}$ and $h=0.05$ both criteria yield 
approximately the same minimum mass, namely $0.4$ \mjup. These criteria
were shown to be valid within a factor of 2 in \cite{1999ApJ...514..344B}.

In our simulations, we found the density reduction factor near the orbit of 
the planet to be more than 100 for a 1 \mjups planet, 10 for a $0.5$ \mjups 
planet and 2 for a $0.1$ \mjups planet. All factors were measured after
$500$ orbits of the planet. In view of these results, we can confirm that
both criteria give a good estimate of the minimum mass for gap opening.
This seems to contradict recent results from \cite{2002ApJ...572..566R}, who
found that it takes only a fraction of $M_\mathrm{min,h}$ to open a gap.
This fraction depends on viscosity and disk mass, and it can be as low as 
$0.1$. However, in that analysis feedback from migration of the planet plays
a role, while our planet remains fixed in the grid. 

\begin{figure}
\resizebox{\hsize}{!}{\includegraphics[bb=35 10 285 245]{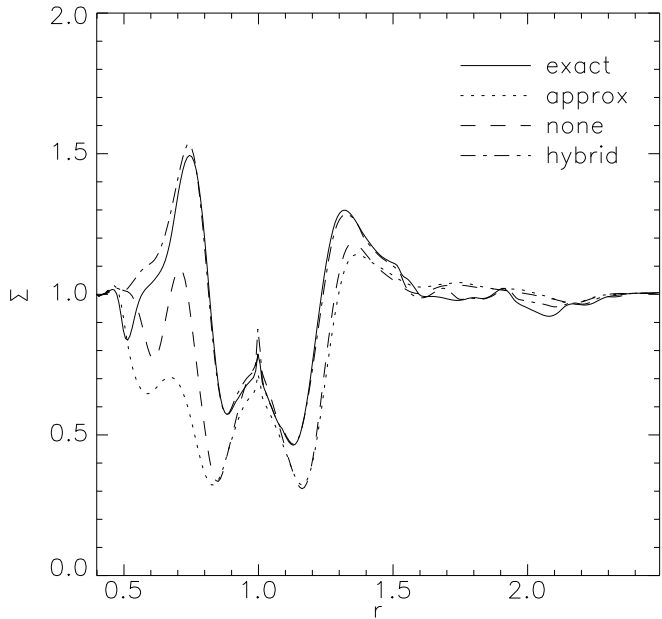}}
\caption{Azimuthally averaged density profile after $20$ orbits for different
implementations of the source term integration: exact stationary extrapolation,
approximate extrapolation and no stationary extrapolation.}
\label{fig4}
\end{figure}

Figure \ref{fig2} (left panel) shows a greyscale plot of the surface
density after $10$ orbits of a 1 \mjups planet, and all the basic features in 
the flow are already visible. We can see two trailing spiral arms 
leaving the planet: the inner arm moves all the way down to the inner 
boundary, while the outer arm reaches to about $r=1.5$. Note also the 
secondary waves excited in the inner and in the outer disk. 

All spiral waves are stationary in the corotating frame. For this run 
we did not take out any matter inside the Roche lobe, so the density reaches 
very high values near the planet.

The formation of the gap is quite a violent proces. There are structures
visible in the corotating region and in the outer edge of the gap. The
physical viscosity is able to damp these fluctuations, leading to a stable
gap after 200 orbits (Fig. \ref{fig2}, right panel). 

The left panel of Fig. \ref{fig3} shows the azimuthally averaged surface 
density for the same run at different times. The gap that is formed is 
approximately $0.5~r_{\rm p}$ wide, with two density bumps on either side. 
After $100$ orbits the inner disk is starting to be accreted onto the central 
star. At all times the position of the planet is visible as a small spike at 
$r=1$. 

The proces of gap formation is numerically challenging, as was already 
shown by \cite{1998A&A...338L..37K}. In the remainder of this section we 
study the formation of gaps as a function of numerical parameters.

\subsection{Source term integration}
\label{secaccSource}
First of all we focus on the integration of source terms. We consider four
different methods. First of all our standard scheme, in which we solve for
the angular momentum exactly, and integrate all other source terms using
Eq. \ref{eqapproxstat}. Secondly, we have the approximate scheme, where all
source terms are integrated using Eq. \ref{eqapproxstat}. Next, we consider
the case of no stationary extrapolation at all (see Eq. \ref{eqHom}), which
is what ordinary Riemann-type schemes would use. Finally, we again integrate 
the angular momentum exactly, and all other source terms using Eq. \ref{eqHom}.
This mimics the fix found by \cite{1998A&A...338L..37K}, and we will refer to
this scheme as \emph{hybrid}, because it combines the two extremes of 
stationary extrapolation and no stationary extrapolation.

In Fig. \ref{fig3}
we show the formation of the gap for the exact stationary extrapolation
(left panel) and the approximate stationary extrapolation (right panel). 
The difference between the two plots is quite dramatic. While both methods
keep angular momentum conserved up to second order, approximate extrapolation
leads to a much wider and deeper gap. After 200 orbits of the planet, even
the region $1.5<r<2.1$ is participating in gap formation. From Fig. 
\ref{fig3} it is clear that approximate extrapolation, while it preserves 
stationary states up to second order, is not the way to go in case of a 
strongly rotating fluid. Only the exact stationary extrapolation produces
results comparable to other codes (see \cite{comparison}).

For these runs the wave damping boundary regions were employed to be able to 
compare to runs with no stationay extrapolation at all. The effect of the 
damping zones is clearly visible after 200 orbits, especially at the inner 
boundary. 

Both models of Fig. \ref{fig3} were also run in an inertial frame for
comparison. For the case of exact extrapolation it made no difference at all,
while only small changes were seen for approximate extrapolation. However,
even in an inertial frame there is a Coriolis force term, see Eq. \ref{eq2D},
so an inertial frame calculation is in this case not a valid check for the 
instability found by \cite{1998A&A...338L..37K}.

\begin{figure}
\resizebox{\hsize}{!}{\includegraphics[bb=35 10 285 245]{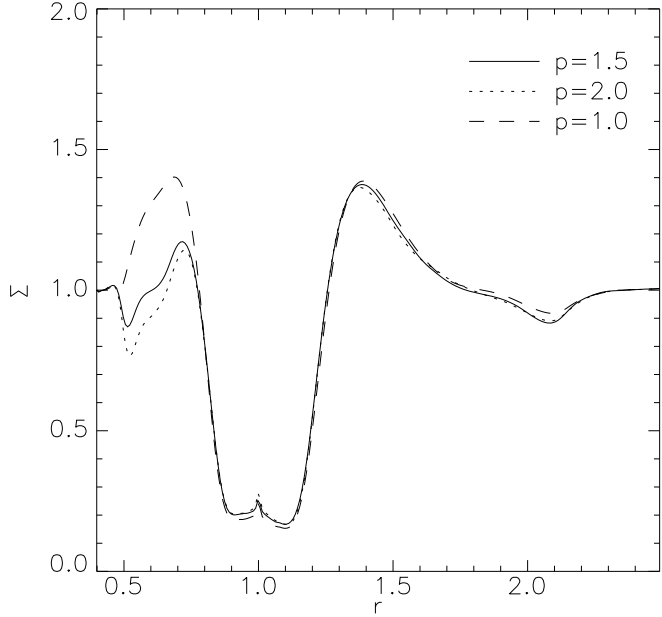}}
\caption{Azimuthally averaged density profile for a 1 \mjups planet 
after $100$ orbits for different flux limiters: $p=1.5$ (standard value), 
$p=2.0$ and $p=1.0$ (see Eq. \ref{eqfluxlim})}
\label{fig5}
\end{figure}

\begin{figure}
\resizebox{\hsize}{!}{\includegraphics[bb=35 10 285 245]{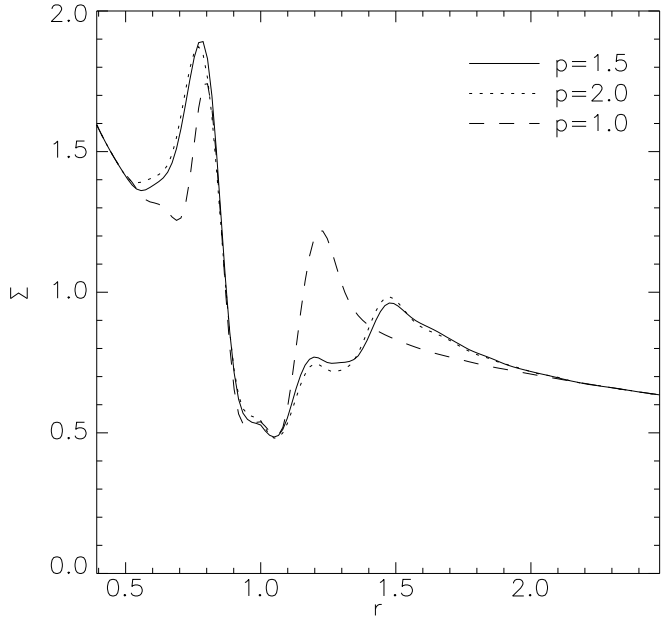}}
\caption{Azimuthally averaged density profile for a $0.1$ \mjups planet 
after $200$ orbits in a non-viscous disk for different 
flux limiters: $p=1.5$ (standard value), $p=2.0$ and $p=1.0$ (see Eq. 
\ref{eqfluxlim})}
\label{fig6}
\end{figure}

\begin{figure*}
\includegraphics[bb=30 10 540 245,width=17cm]{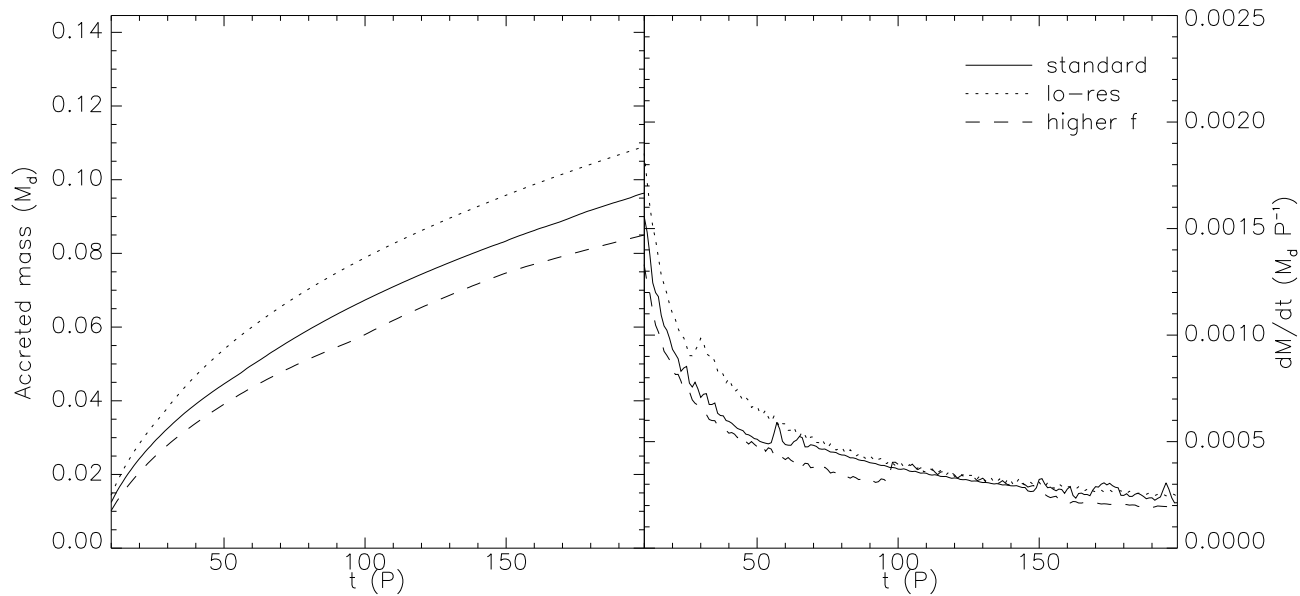}
\caption{Accretion onto a 1 \mjups planet. Left panel: total accreted mass
in units of the initial disk mass. Right panel: accretion rate, in units of
disk masses per orbit.}
\label{fig7}
\end{figure*}

In Fig. \ref{fig4} we compare the azimuthally averaged surface density 
profile for the four methods of source integration: standard, approximate, no
stationary extrapolation at all, and the hybrid form. This figure shows that 
both the approximate extrapolation and the case of no stationary extrapolation 
overestimate the angular momentum transport, which leads to spurious 
gap widening. It is remarkable, however, that approximate extrapolation does 
the worst job, worse even than the run without stationary extrapolation. This 
is especially clear for the inner disk ($r<1$). It is interesting to note that 
in this case trying to deal with the issue raised by \cite{1998A&A...338L..37K}
in an approximate way actually makes matters worse than for the standard 
source term integration.

The hybrid extrapolation reproduces the correct gap width and gap depth 
after $20$ orbits, and follows closely the exact extrapolation almost 
everywhere. Only in regions where the strongest waves exist, near the planet
and in the inner disk, the two methods give different results. Qualitatively
hybrid extrapolation gives similar results as a more diffusive flux limiter
(see Sec. \ref{secaccflux}).

We also performed the same simulations at a lower resolution. For the case 
of exact and hybrid extrapolation no differences were found, but for both 
other cases the width and structure of the gap changed significantly, 
especially for the approximate stationary extrapolation. It is clear from Eq. 
\ref{eqapproxstat} that the error made in this approximate scheme will 
depend on the spatial resolution of the grid. On a grid of only
$(n_\mathrm{r},n_\mathrm{\phi})=(128,384)$ the effect of gap widening is
even more dramatic than seen in Fig. \ref{fig3}. The inner disk is cleared 
away even faster, leading to an unstable situation near the boundary. For
the case of no stationary extrapolation the effect is not that severe, but
still the gap is wider than in Fig. \ref{fig4}. 

\subsection{Flux limiter}
\label{secaccflux}
The flux limiter is basically a switch between using a first or second order
interface flux for the state update. Near shocks it should be first order,
in smooth flow it should be second order. Applying a second order flux near
a discontinuity leads to numerical smearing of the state, and therefore 
the flux limiter that switches the fastest to first order fluxes gives 
the least numerical diffusion. Here we study the effect of this numerical
diffusion on the formation and appearance of the gap.

In Fig. \ref{fig5} we show the gap structure for three different limiters 
(see Eq. \ref{eqfluxlim}): $p=1.0$ (minmod, most diffusive), $p=2.0$ (superbee,
least diffusive) and $p=1.5$ which we call soft superbee. In the outer disk
all limiters give more or less the same result. But in regions where the
waves induced by the planet are the strongest, the inner disk and close to 
the planet, we see clear differences. The diffusive minmod limiter damps
the ingoing waves more than the other limiters, leading to an enhanced 
surface density near $r=0.7$. Close to the planet this higher diffusion 
also leads to an enhanced surface density, seen as a spike at $r=1$ in
Fig. \ref{fig5}.

There are no large differences in the results obtained with the superbee
limiter and its softer version. This indicates that our choice of $p=1.5$
(see Sec. \ref{secfluxlim}) is a good trade-off between the two extremes
minmod and superbee. It inherits the low numerical diffusion from superbee,
as is clear from Fig. \ref{fig5}, while it is more stable against numerical
overshoots.

Since the different limiters imply different numerical viscosities, it 
is interesting to study how they influence the result in the case when there 
is no physical viscosity added to the simulations. Then the situation changes 
drastically, because numerical viscosity starts to play a major role, in
particular the length over which shocks are damped. As an illustration, 
we show the azimuthally averaged surface density for a $0.1$ \mjups planet 
in a viscosity-free disk in Fig. \ref{fig6}, for the three different flux 
limiters. It is clear that for the diffuse minmod limiter the waves are
damped much faster, and the angular momentum is deposited much closer to the 
planet. For the superbee limiter, as well as its softer version, a much 
wider gap results. 

\section{Accretion rates}
\label{secAcc}
Now we turn to the problem of accretion onto the planet. The growth rate
of the planet is important because it determines the ultimate mass the 
planet will reach. Two-dimensional studies of \citet{2002A&A...385..647D}
and \citet{1999ApJ...526.1001L} showed that a planet of $1$ $\mathrm{M_J}$
grows approximately at a rate of $10^{-4}$ $\mathrm{M_d P^{-1}}$, where
$\mathrm{M_d}$ is the disk mass within the computational domain. However,
the study by \cite{1999MNRAS.303..696K} done at lower resolution indicated 
an accretion rate more than ten times higher.

In this section we look at three different mass regimes: high-mass planets,
which open clear gaps in the disk, low-mass planets, which do not open gaps,
and intermediate mass planets, which create only small density dips around 
their orbit.

\begin{figure}
\resizebox{\hsize}{!}{\includegraphics[bb=154 290 435 545]{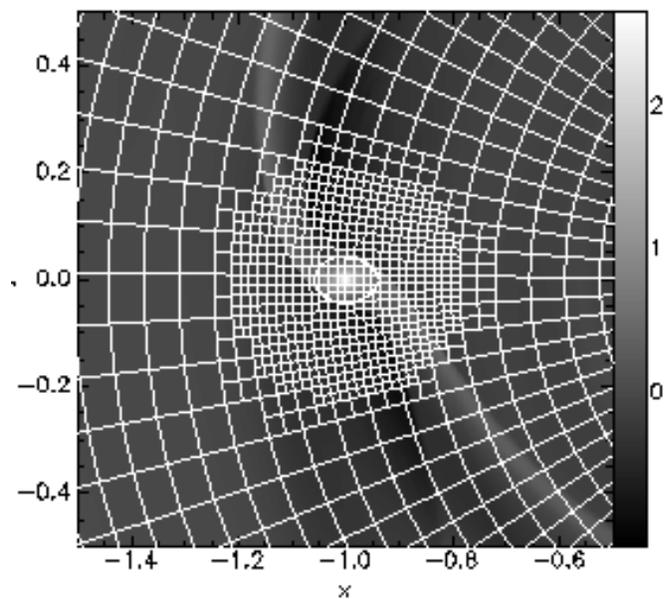}}
\caption{Close-up on the logarithm of the surface density near the planet. 
Overplotted is the AMR mesh structure, where each block consists of 8x8 cells.}
\label{fig8}
\end{figure}

\begin{figure}
\resizebox{\hsize}{!}{\includegraphics[bb=15 10 285 240]{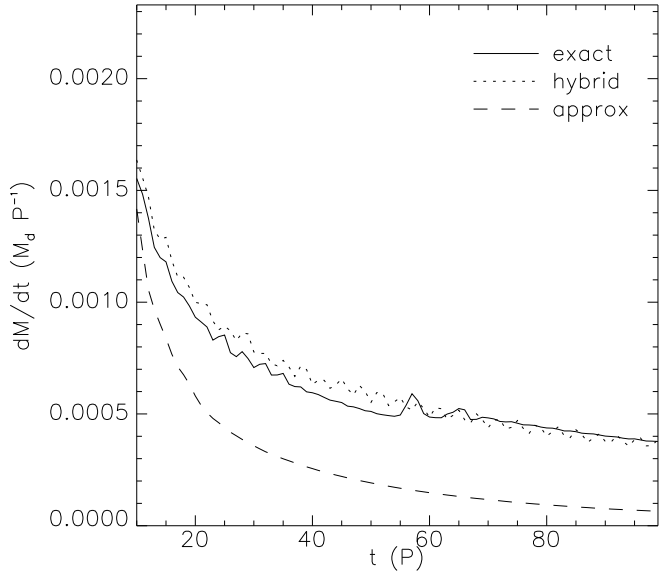}}
\caption{Accretion rates onto a 1 \mjups planet for three different methods
of source term integration.}
\label{fig9}
\end{figure}

\begin{figure}
\resizebox{\hsize}{!}{\includegraphics[bb=15 10 285 240]{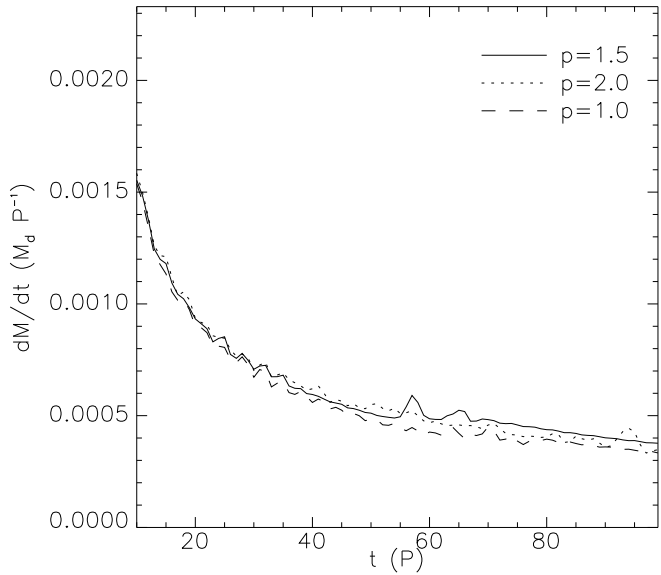}}
\caption{Accretion rates onto a 1 \mjups planet for three different flux
limiters.}
\label{fig10}
\end{figure}

\subsection{High mass planet}
We start by discussing the results for a $1$ \mjups planet. Because this
planet is able to open up a gap in the disk, the accretion rate is determined
by the amount of mass that flows through the gap \citep{1999MNRAS.303..696K},
and less by the density structure close to the planet. 

In Fig. \ref{fig7} we show the accreted mass and the accretion rate as a 
function of time for our standard resolution, accretion parameters, source
term integration and flux limiter (solid lines). Because we do not start with
an initial gap the accretion rate is very high at the beginning of the 
simulation, dropping about one order of magnitude in $200$ orbits to a value
of $2~10^{-4}$ (disk masses per orbit). We ran one model to $500$ orbits, and
the final accretion rate turned out to be  $1.5~10^{-4}$. 
The fact that the accretion rate approaches a constant value after about 500 
orbits and the actual value found agree within a factor 2 with 
\citet{2002A&A...385..647D} and \citet{1999ApJ...526.1001L}. The planet 
accretes approximately 10 percent of the total disk mass during the simulation.

\subsubsection{Accretion parameters}
The accretion procedure is described by two parameters: the radius within
which we take out material ($r_\mathrm{acc}$) and the value of $f$ (see Eq. 
\ref{eqaccproc}). We can vary these to see if this influences the final
accretion rate.

In Fig. \ref{fig7} the accretion rates onto a 1 \mjups planet are shown 
for the standard case ($r_\mathrm{acc}=0.5$ $\rm{R_R}$ and $f=0.5$) and for 
parameters $r_\mathrm{acc}=0.1$ $\rm{R_R}$ and $f=5/3$ (the curve labeled with 
'higher f'). The standard set
was used by \cite{1999MNRAS.303..696K}, and the other case by 
\cite{2002A&A...385..647D}. Note that the accretion areas differ by a factor
$25$, while $f$ varies only by a factor $3.3$. So for identical density 
distributions close to the planet the standard parameters yield $7.5$ times
more accreted mass during one time step. Despite this the final accretion
rate is the same for both sets of parameters.

Because of the different accretion radii any difference in accretion rate
would imply that the flow within the Roche lobe is important for the accretion
proces. Disk material that makes it to $0.5$ $\rm{R_R}$ is accreted for the
standard parameters, while it has to make it all the way to $0.1$ $\rm{R_R}$
to be accreted in the second case. The fact that we do not see any differences
indicates that the accretion rate is determined by the amount of mass the 
disk is able to supply, independent of local processes near the planet. 

\begin{figure*}
\includegraphics[bb=30 10 540 245,width=17cm]{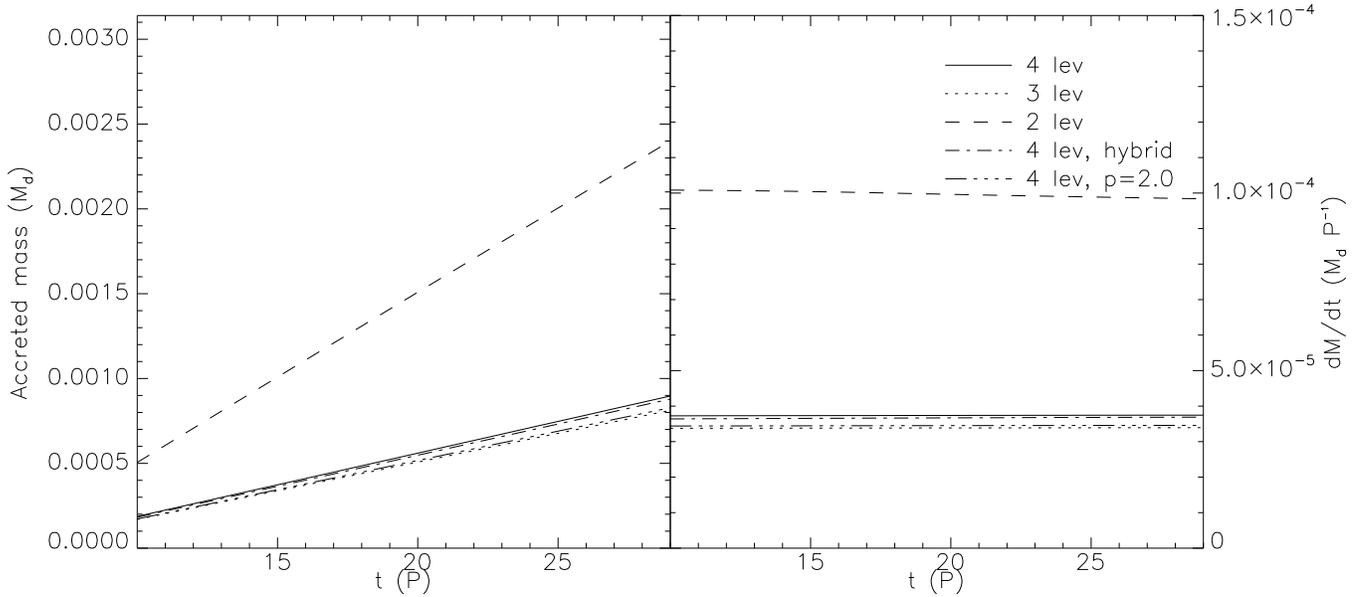}
\caption{Accretion onto a 1/64 \mjups planet. Left panel: total accreted mass
in units of the initial disk mass. Right panel: accretion rate, in units of
disk masses per orbit.}
\label{fig11}
\end{figure*}

\subsubsection{Resolution}
\label{secAccres}
Our base resolution corresponds to approximately 8 cells per $\rm{R_R}$, which
means that the standard accretion area is resolved by only a few grid cells.
This might be of influence on the inferred accretion rate, and therefore
we performed simulations on different resolutions.

First of all we lowered the resolution with a factor of 2, the same resolution
as in \cite{1999MNRAS.303..696K}. In this case we do not resolve the Roche
lobe, so if the flow close to the planet is important for accretion we would
expect differences. However, since the two sets of accretion parameters did 
not yield different accretion rates we expect no big resolution effects. 
Indeed, Fig. \ref{fig7} shows that for this low resolution case (dotted line)
the accretion rate is the same as for our standard resolution. This shows that
resolving the flow within the Roche lobe is not important for gap-opening 
planets, at least not for determining accretion rates.

It is computationally very expensive to go up a factor of 2 in resolution
on the whole grid, therefore we used our AMR module to refine the region
around the planet. Figure \ref{fig8} shows a close-up on the density pattern 
after 20 orbital periods. Overplotted are the Roche lobe and the grid 
structure. Each block represents 8x8 cells, so that we have approximately 
1500 cells within the Roche lobe. Therefore we can really resolve the flow 
inside the Roche lobe with this resolution. Nevertheless also this resolution
yielded the same accretion rate of $2~10^{-4}$ $\rm{M_d} P^{-1}$. 

\subsubsection{Equation of state}
It is interesting to compare accretion rates for a truely isothermal
simulation and a run that does include an energy equation, but at a very
low adiabatic exponent $\Gamma$. This has been done before to mimic 
isothermal flow for planet-disk interaction \citep{2003ApJ...589..556N}.
The basic idea is that for a low value of $\Gamma$ the gas can be 
compressed without a large change in temperature. In order to check the 
validity of this approach regarding planet-disk interaction, we ran 
simulations including an energy equation but with a low value of $\Gamma$.

We performed two simulations including an energy equation, one with 
$\Gamma=1.001$ and one with $\Gamma=1.01$. The latter value was also used
by \citet{2003ApJ...589..556N}. We found that the basic flow structures 
remain the same, and that the temperature profile does not change anywhere
but very close to the planet. For $\Gamma=1.001$ the temperature rises 
already by a factor 10, and for $\Gamma=1.01$ by a factor of $60$. This
steep temperature gradient slows down the gas flow towards the planet 
considerably.

Due to the higher temperatures close to the planet the accretion slows 
down. Already for $\Gamma=1.001$ the accretion rate drops by a factor of 
$2$, and even a factor of $10$ for $\Gamma=1.01$. This shows that the 
accretion rate
depends very much on the temperature near the planet, and that 'nearly'
isothermal simulations can produce very different results from truely
isothermal runs.

This effect is different from the one described in 
\cite{1999MNRAS.303..696K}, who found that a polytropic equation of state 
leads to a reduction in the accretion rate. Because the temperature 
is proportional to the density in that case (for $\Gamma=2$), the gap
region is much cooler and therefore the viscosity is reduced when the
$\alpha$-formalism is used. In our case, there is a temperature \emph{rise}
very close to the planet, which leads to a pressure barrier that is able
to slow down accretion considerably, even when $\Gamma$ is as low as
$1.001$. Therefore we conclude that these kinds of simulations are not able
to mimic isothermal flow for this specific case, due to the deep potential
well of the planet.

\subsubsection{Source term integration}
\label{secAccSource}
In Fig. \ref{fig3} we demonstrated that the way source terms are 
included has dramatic effects on the proces of gap formation. In this 
section we show that this also affects the accretion onto the planet.

Figure \ref{fig9} shows the accretion rates for three different methods
of source term integration: exact, hybrid and approximate extrapolation.
See Sec. \ref{secGap} for their definition. Again, as in Fig. \ref{fig3} 
approximate extrapolation is the odd one out, yielding a 4 times lower
accretion rate. This clearly has to do with the difference in gap formation
time scale, and again it is clear that approximate extrapolation is not the 
way to go.

Both the exact and hybrid method give the same results. This is not surprising,
because the region where the methods differ most is where the source terms 
are large: close to the planet. And as we mentioned before, this region is
not important for the accretion rate.  

\begin{figure*}
\includegraphics[bb=30 10 540 245,width=17cm]{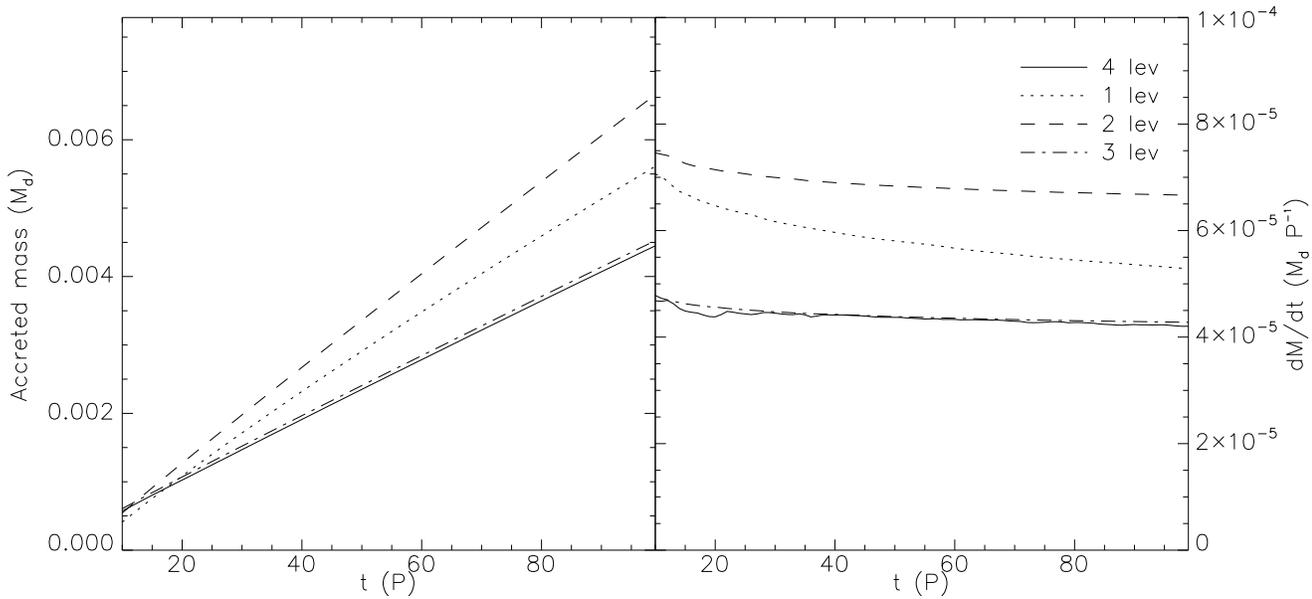}
\caption{Accretion onto a 1/8 \mjups planet. Left panel: total accreted mass
in units of the initial disk mass. Right panel: accretion rate, in units of
disk masses per orbit.}
\label{fig12}
\end{figure*}

\subsubsection{Flux Limiter}
\label{secAccFlux}
The flux limiter did not play a major role in gap formation (see Fig. 
\ref{fig5}), only at the inner parts of the disk some differences can be seen.
Figure \ref{fig10} shows the accretion rates for the three different
limiters discussed in Sec. \ref{secGap}. It is clear that the accretion 
rate does not depend at all on the flux limiter. 

This can be understood if one realizes that it is the flow from the outer disk
to the inner disk that governs the accretion rate (see 
\cite{1999MNRAS.303..696K}). But in the outer disk the waves are weaker than
in the inner disk, so a different flux limiter should not change the mass
flux from the outer disk very much.

\subsection{Low mass planet}
\label{secAccLow}
We now move to the other side of the planetary mass spectrum to investigate
accretion onto planets that do not open gaps in the disk. Specifically we
focus on a planet with mass $\rm{M_J}/64$. Because 
$\rm{R_R} \propto M_{\rm{p}}^{1/3}$ the Roche lobe of this planet is $4$ times
as small as the Roche lobe of a jupiter-mass planet.

In Fig. \ref{fig11} the solid line gives the accretion rate for our standard
parameters and 4 levels of AMR. For our base resolution the Roche lobe would 
only be resolved by only 1 grid cell, clearly not enough to study accretion.
Figure \ref{fig11}, dashed line, shows that 2 levels of refinement, yielding 
4 cells per $\rm{R_R}$, is still not enough to reproduce the result for 4
levels of AMR. The accretion rates for 3 and 4 levels of AMR agree very well,
showing that at least 8 cells per $\rm{R_R}$ are needed for accurately 
modeling accretion. 

All accretion rates have reached their final value after $30$ orbits. The 
model with 4 levels of refinement was run until 200 orbits with no change
in the accretion rate. This is because the planet does not open up a gap,
which would take about $200$ orbits (see Sec. \ref{secGap}). The value of
$3.5~10^{-5}$ that we find is in good agreement with the results of
\cite{2002A&A...385..647D}. Note however that they need 7 grid levels,
corresponding to approximately 6 levels of AMR for our simulations, while
we need only 3 levels to obtain the same result.

Because we have already discarded the approximate extrapolation scheme in
Sec. \ref{secGap} and Sec. \ref{secaccSource} we looked only at the difference
between exact extrapolation and hybrid extrapolation for the low-mass case.
However, we point out that in this case approximate extrapolation gave 
identical results. This is because the waves from this planet are too weak
to make a difference in angular momentum flow. From Fig. \ref{fig11} we see 
that both methods we considered for source term integration
give identical results. In this case the planetary gravitational source terms
are too small to cause a large difference in accretion rate.

Also different flux limiters gave identical results. In Fig. \ref{fig11} we
only show the superbee-result, but the minmod limiter yielded exactly the
same accretion rate. This was to be expected, because in the limit of smooth
flow (or, equivalently, no strong waves, as is the case for a low-mass planet)
all limiters produce the same flux (see Eq. \ref{eqfluxlim} with $a_0 = a_u$).

\begin{figure}
\resizebox{\hsize}{!}{\includegraphics[bb=15 10 285 240]{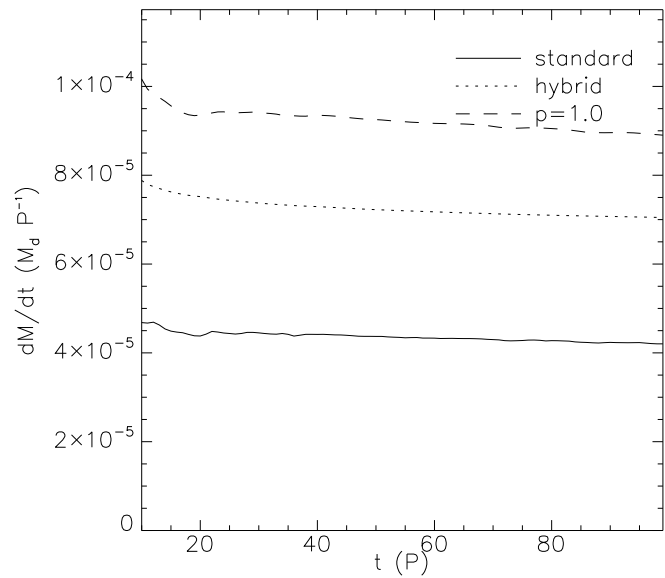}}
\caption{Accretion rates onto a 1/8 \mjups planet for the standard case 
(solid line), hybrid extrapolation (dotted line) and the minmod limiter
(dashed line).}
\label{fig13}
\end{figure}

\begin{figure*}
\includegraphics[bb=10 10 530 248,width=17cm]{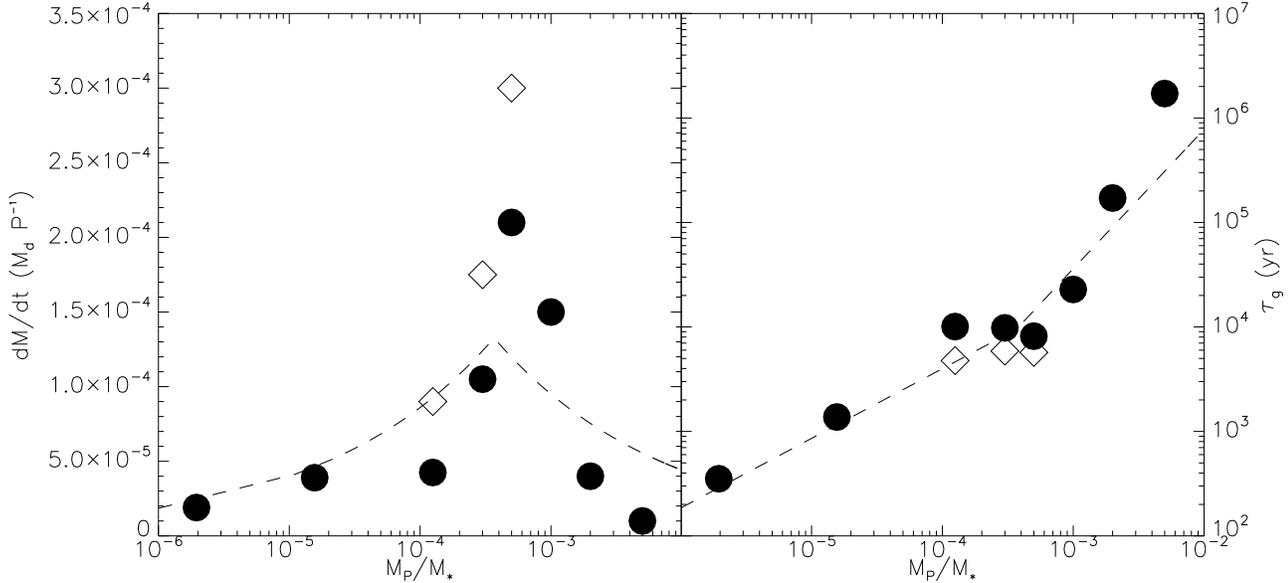}
\caption{Left panel: accretion as a function of planet mass. Right panel:
growth time (see Eq. \ref{eqgrow}). Filled circles indicate our standard 
result, diamonds give the results for the diffusive minmod limiter wherever
they were different. In both panels the fit of \cite{2002A&A...385..647D} 
is shown by the dashed line.}
\label{fig14}
\end{figure*}

\subsection{Intermediate mass planet}
\label{secAccMed}
The masses that lie in between the two extremes we have considered so far
are perhaps the most interesting. In this regime the transition from 
linear disk response to gap formation takes place, and therefore also the
transition from Type I to Type II migration. \cite{2002A&A...385..647D} find
the highest accretion and migration rates here. Also, the cores of giant
planets may well be in this mass range.

For this section we focus on a planet of mass $1/8$ \mjup, which makes its
Roche lobe twice as small as for a Jupiter-sized planet. In Fig. \ref{fig12} 
we compare the accretion rates for different resolutions. First of all, note
that we need a relatively high resolution to obtain convergence (3 levels of
refinement, which amounts to 16 cells per $\rm{R_R}$). This is already an 
indication that interesting things are going on. Secondly, we find an accretion
rate that is about twice as low as was found by \cite{2002A&A...385..647D}.
In view of the good agreement for the low-mass planet this is remarkable.

In Fig. \ref{fig13} we compare the standard model with the method of hybrid
extrapolation and the minmod flux limiter. While for the high-mass planet
as well as the low-mass planet these three methods gave identical results,
for a $1/8$ \mjups planet they differ significantly, giving a $66~\%$ and
$100~\%$ higher accretion rate for the hybrid extrapolation and minmod
limiter, respectively. With the most diffusive flux limiter we can 
reproduce the result of \cite{2002A&A...385..647D}. This shows that numerical
diffusion is a very important issue in these kinds of simulations.

It is not surprising that the diffusive minmod limiter increases the accretion
rate, because it tends to smear out the strong density and velocity gradients
near the planet, allowing mass to diffuse into the Roche lobe. For this planet
local conditions determine the accretion rate, unlike the previous cases. 
Low-mass planets do not excite strong enough waves to alter their environment
significantly, while high-mass planets open up gaps, and the accretion rate
is therefore determined by the global evolution of the disk. Planets of
intermediate mass, however, do not clear a gap while they excite reasonably
strong waves, making the dynamics very interesting. 

A similar story applies to the source term integration. We have seen no
differences between exact and hybrid extrapolation for high and low-mass
planets, while for our $1/8$ \mjups planet the difference is quite dramatic.
Again, the gravitational forces due to low-mass planets are too weak to
cause a difference, while for high-mass planets the local conditions are
irrelevant. 

\subsection{Dependence on planetary mass}
In the previous sections we have looked in detail at three characteristic
planetary masses. Here we focus on how the accretion rate depends on the
mass of the planet. We have run simulations for planets from $0.5$ 
$\rm{M_\oplus}$ up to $8$ \mjup, or, in other words, from deep in the 
linear regime to well above the gap-opening mass.

In the left panel of Fig. \ref{fig14} we show the accretion rate for all 
planetary masses, as well as the relation found by \cite{2002A&A...385..647D}.
As was mentioned in Sec. \ref{secAccLow} the results for the low-mass planets
agree very well with those found by \cite{2002A&A...385..647D}. However, as 
soon as the disk response to the planet approaches the non-linear regime
around $M_p \approx 0.1$ \mjups the results start to differ significantly.
At first, the accretion rate stays low, but around $M_p \approx 0.2$ \mjups
there is a strong rise leading to a maximum at $0.5$ \mjup, followed by a 
steep decline. 

The general features of the plot are consistent with the results of 
\cite{2002A&A...385..647D}: the accretion rate rises with planetary mass, with 
a maximum around $0.5$ \mjup, followed by a steeper decline. However, the rise
as well as the decline are much more dramatic in our case, leading to a
higher maximum accretion rate and lower accretion rates for the highest mass
planets.

The diamonds in Fig. \ref{fig14} represent the results obtained with the 
diffusive minmod limiter wherever they differed significantly from the 
standard case. We conclude that a diffusive flux limiter always tends to
increase accretion onto the planet, especially in the mass regime in 
which the transition from linear to non-linear disk response takes place. 

The difference between the results obtained with the minmod flux limiter and 
our standard limiter can be as large as $50 \%$ (see Fig. \ref{fig14}). 
Because these differences are caused by purely numerical effects, one can 
interpret these as an estimate of the error in the values of the accretion
rates in this mass regime. Keeping these error estimates in mind, we see that
our results are roughly in agreement with the relation found by 
\cite{2002A&A...385..647D}. Note, however, that our standard results always
represent the lowest possible accretion rate. Different flux limiters or
different source term integrations always lead to a higher accretion rate.  

The growth time scale is defined as the time it takes for a planet to 
double its mass at a given accretion rate:
\begin{equation}
\label{eqgrow}
\tau_{\rm g} \equiv \frac{M_{\rm{p}}}{dM_{\rm{p}}/dt}
\end{equation}
In order to express $\tau_{\rm g}$ in years we took a disk mass of $0.0035$
\msun. When for a given mass $\tau_{\rm g}$ becomes comparable to the total 
disk life time the planet can not grow beyond this mass. This limiting mass 
is of the order of several \mjups \citep{2002A&A...385..647D}. However, they
do not consider planets more massive than $1$ \mjup. 

In the right panel of Fig. \ref{fig14} we show $\tau_{\rm g}$ for all planet
masses. Again, the dashed curve shows the fit from \cite{2002A&A...385..647D}.
Because of the logarithmic scale on the y-axis the differences are much less
pronounced. However, we still see the two different regimes and the sharp 
transition. Very important is the steep rise in $\tau_{\rm g}$ for the 
highest masses. Assuming that the planet spends maximal $10^6$ years inside
this disk the maximum mass it can reach is $4$ \mjup. Note that if we exclude
the planets more massive than $1$ \mjups the slope of the growth time scale
is in excellent agreement with the relation found by 
\cite{2002A&A...385..647D}. Extrapolating this slope we would find that the 
maximum mass that can be reached is about $10$ \mjup. This shows that it
is really necessary to include the highest mass planets to obtain a reliable
estimate of the maximum planetary mass that can be reached through disk 
accretion. 

The decline in accretion rate that we find for high-mass planets is 
steeper than found by \cite{1999ApJ...526.1001L}, who used the ZEUS code
for their simulations. We have shown that this is not due to the flux 
limiter or the source term integration, and the reason for this 
difference may be found in the intrinsic difference between the 
finite-difference approach and Riemann solvers. However, this is impossible 
to verify within our numerical method.

\begin{figure}
\resizebox{\hsize}{!}{\includegraphics[bb=213 10 505 245]{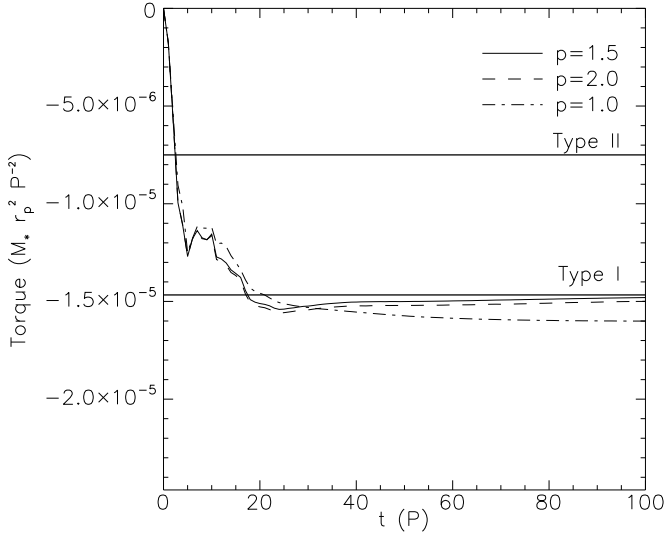}}
\caption{Torque exerted on a $1/8$ \mjups planet, for three different flux 
limiters. The whole Roche lobe is excluded for these calculations.}
\label{fig15}
\end{figure}

\begin{figure}
\resizebox{\hsize}{!}{\includegraphics[bb=213 10 505 245]{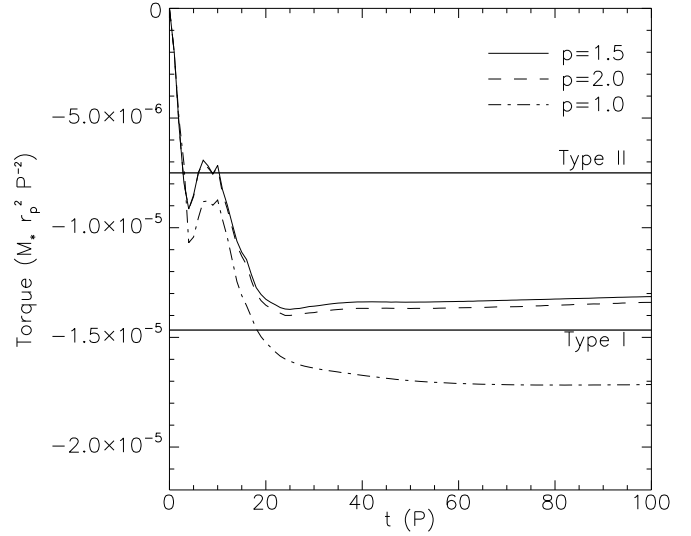}}
\caption{Torque exerted on a $1/8$ \mjups planet, for three different flux 
limiters. Only the inner half of the Roche lobe is excluded for these 
calculations.}
\label{fig16}
\end{figure}

\section{Migration}
\label{secMig}
It has become clear in recent years that protoplanets can be extremely mobile
within their protoplanetary disk. Three types of migration can be 
distinguished: Type I, which concerns low-mass planets that do not open gaps
in the disk , Type II for migration inside a clear gap, and Type III, for
which the radial movement of the planet drives its migration 
\citep{2003ApJ...588..494M}. Because our planet stays at a fixed location,
we are only dealing with Type I and Type II migration. 

For the torque calculations, we exclude material orbiting within the Roche 
lobe of the planet, and we assume an initial disk mass of $0.0035$ \msun. 

\subsection{Numerical Parameters}
Migration rates turn out to be very robust once the Roche lobe as well as the
disk scale height is well resolved. Across the whole mass spectrum we found no
significant torque differences for the various methods of source term 
integration and flux limiters. Even approximate extrapolation, which was shown
to lead to spurious numerical evolution of the gap, does not affect the torque
on the planet. This is because for low-mass planets approximate extrapolation 
does not alter the density structure around the planet (see Sec. 
\ref{secAccLow}), and for high-mass planets it only speeds up gap formation,
which leads to Type II migration. 

As an example we show in Fig. \ref{fig15} the torque as a function of time for 
a $1/8$ \mjups planet for different flux limeters. All three limiters lead to
a torque that is comparable to Type I migration. As with gap formation, the 
difference between $p=1.5$ and $p=1.0$ is larger than the difference between
$p=1.5$ and $p=2.0$ (see Fig. \ref{fig5}), but for the torque the effect is
not significant. This shows that the sensitivity of the accretion rate on
numerical parameters found in Sec. \ref{secAccMed} is due to effects 
{\it inside} the Roche Lobe, while we explicitly excluded this region for the
torque calculations of Fig. \ref{fig15}. 

This is further illustrated in Fig. \ref{fig16}, where we again show the total
torque on the planet, but now we exclude only the inner half of the Roche 
lobe. The first thing to note is that the minmod limiter now gives a
significantly different result compared to the other two. We found similar
behaviour for the accretion rate in Sec. \ref{secAccMed}.

Secondly, we observe that the torque is less negative for 
$p=1.5$ and $p=1.0$. This is the torque reversal also observed by 
\cite{2002A&A...385..647D}: material within the Roche lobe exerts a 
{\it positive} torque on the planet, slowing down migration. However, it is
not clear exactly where the disk ends and the envelope of the planet starts.
A gas giant is usually assumed to fill its Roche lobe during formation
\citep{1996Icar..124...62P}, so all material orbiting inside the Roche lobe
is part of the envelope. Therefore the question rises whether a planet should
change its orbit due to its own envelope. On the other hand, important 
physical processes are not included in the current hydrodynamical simulations
of planet-disk interaction, namely heat transport (due to the isothermal 
assumption) and self-gravity. Both can have dramatic influence on the direct
surroundings of the planet, where the density is highest and the temperature 
may differ significantly from the ambient disk, and therefore it is not 
feasible yet to make a connection between the one-dimensional planet formation
models of \cite{1996Icar..124...62P} and the hydrodynamic models of planet-disk
interaction. For now, we are only interested in comparing our values to 
previous analytical and numerical results, which will be done in the next 
section.

\begin{figure}
\resizebox{\hsize}{!}{\includegraphics[bb=247 10 505 240]{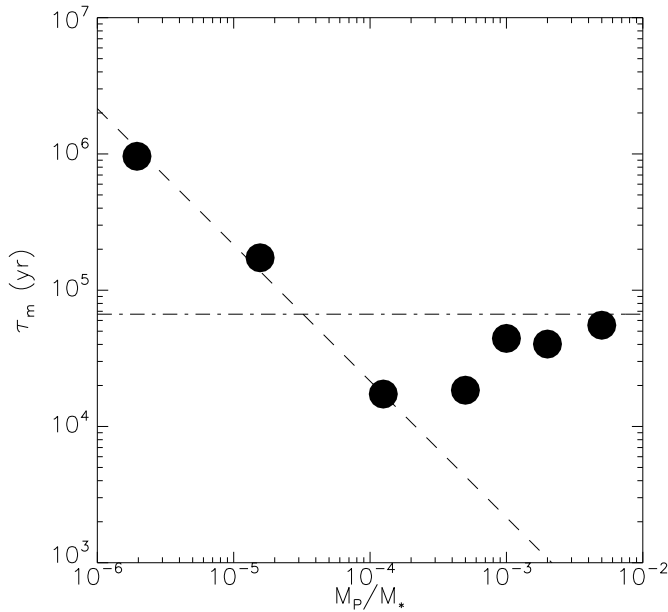}}
\caption{Migration time scales for planets of different mass. Dashed line:
analytical solution by \cite{2002ApJ...565.1257T} for Type I migration. 
Dash-dotted line: result by \cite{1997Icar..126..261W} for Type II migration.}
\label{fig17}
\end{figure}

\subsection{Dependence on planetary mass}
For the low-mass case of Type I migration there exists a nice analytical
prediction of the migration rate as a function of planetary mass 
\citep{2002ApJ...565.1257T}. For gap-opening planets, the migration rate 
should not depend on planetary mass, only on the viscosity in the disk 
\citep{1997Icar..126..261W}. Numerical simulations show that in between these 
two extremes the fastest migration takes place, in two-dimensional simulations
\citep{2002A&A...385..647D} as well as in three-dimensional simulations
(\citet{2003MNRAS.341..213B,2003ApJ...586..540D}).

In Fig. \ref{fig17} we show the migration time scale $\tau_{\rm m}$ defined
by:
\begin{equation}
\tau_{\rm m} = \frac{r_{\rm p}}{\left| v_{\rm r,p} \right|}
\end{equation}
where the radial velocity of the planet due to a torque $\mathcal T$ is given
by:
\begin{equation}
v_{\rm r,p}=2 r_{\rm p} \frac{\mathcal T}{L_{\rm p}}
\end{equation}
Here $\mathcal T$ is the torque due to the disk {\it on} the planet. 

From Fig. \ref{fig17} we see that we can reproduce Type I as well as Type II
migration. The transition region extends approximately from $M_{\rm p}=0.1$
\mjups to $M_{\rm p}=1.0$ \mjup, consistent with the three-dimensional
results of \cite{2003MNRAS.341..213B}. A planet of mass $0.1$ \mjups has
the fastest migration time scale of approximately $10^4$ years.

Comparing Figs. \ref{fig14} and \ref{fig17} we see that for the low-mass 
planets the growth time scale is much shorter than the migration time scale,
while for the high-mass planets it is the other way around. In the 
intermediate case, both time scales are approximately equal.

\section{Summary and conclusion}
\label{secSum}
We have presented a new method for modeling disk-planet
interaction. Key features of the RODEO method are: it is a conservative
method, it treats shocks and discontinuities correctly, and it uses
stationary extrapolation to integrate the geometrical and
gravitational source terms.

We found that the RODEO method performs very well on the problem
of disk-planet interaction, not in the least because we do not
experience serious computational difficulties as for example in
\citet{1998A&A...338L..37K}. We have shed some new light on this matter
in that this instability can be seen to result from the failure of 
most hydrodynamic schemes to recognize stationary states. 

We find that the proces of gap formation crucially depends on source term
integration. Only exact and hybrid extrapolation produce gaps of correct
width. A diffusive flux limiter leads to a more pronounced inner edge of
the gap. 
 
The accretion rates turn out to be very robust in the low-mass regime
($M_{\rm p} < 0.1$ \mjup) as well as in the high-mass regime 
($M_{\rm p} >= 1.0$  \mjup): they are independent of
numerical resolution, accretion parameters and source term integration. 
A different equation of state affects the accretion rates significantly:
even a very low $\Gamma$ of $1.001$ that is used frequently to mimic
isothermal flow the accretion rate drops by a factor of $2$.

For intermediate mass planets the accretion rates are far less certain. 
They are dependent on the flux limiter used as well as on the way source 
terms are integrated. Within our method we find differences of $50 \%$,
and the differences between our results and the similar study of 
\cite{2002A&A...385..647D} are of the same order. This shows that the error
bars on these values are very large.
 
For the highest mass planets we find significantly lower accretion rates 
than in previous numerical studies, limiting the maximum planet mass that 
can be reached to about 4 \mjup. Note that this is in fact the first numerical
study that consistently goes beyond this limiting mass so that no extrapolation
is required.

Migration rates are far more robust than accretion rates, as long as we limit
the disk region that exerts a torque on the planet to outside the planetary
Roche lobe. Therefore the region where most differences arise in the accretion
rates is located deep within the Roche lobe. 

We can nicely reproduce the analytical results for Type I and Type II 
migration, with a transition region extending from $M_{\rm p}=0.1$ \mjups to 
$M_{\rm p}=1.0$ \mjup. In this transition region the fastest migration rate
corresponds to a mass of $0.1$ \mjup. 

The RODEO method can also be used for two-fluid calculations (see 
\cite{2004A&A...425L...9P}) to model dust-gas interaction in protoplanetary 
disks. Also, the method can easily be extended to three dimensions, and is
also suited to treat an energy equation in a multi-dimensional set-up. 
We will consider these extensions in forthcoming papers.

\begin{acknowledgements}
The authors would like to thank Pawel Artymowicz for stimulating
discussions and useful suggestions.  The research of GM has been made
possible by a fellowship of the Royal Netherlands Academy of Arts and
Sciences. This work was sponsored by the National Computing Foundation
(NCF) for the use of supercomputer facilities, with financial support
from the Netherlands Organization for Scientific Research (NWO).
\end{acknowledgements}

\appendix
\section{Explicit expressions}
\label{appA}
Here we give explicit expressions of the eigenvalues, eigenvectors and
projection coefficients for the isothermal Euler equations in
cylindrical coordinates.

\subsection{Radial direction}
The Jacobian matrix $\mathcal{A}$ is given by:
\begin{equation}
\mathcal{A}=\left( \begin{array}{ccc} 0 & 1 & 0 \\ c_s^2-v_r^2 & 2 v_r
& 0 \\ -v_r v_\phi & v_\phi & v_r
\end{array} \right)
\end{equation}
The eigenvalues of this matrix are:
\begin{eqnarray}
\lambda_1 &=& v_r - c_s \nonumber\\ 
\lambda_2 &=& v_r + c_s \\ 
\lambda_3 &=& v_r~~~~~~ \nonumber
\end{eqnarray}
The corresponding eigenvectors:
\begin{eqnarray}
\vec{e}_1 & = & \left(1,v_r-c_s,v_\phi \right) \nonumber \\ 
\vec{e}_2 & = & \left(1,v_r+c_s,v_\phi \right) \\ 
\vec{e}_3 & = & \left(0,0,1\right)~~~~~~~~~ \nonumber
\end{eqnarray}
A vector $\vec{\Delta}\equiv(\Delta_\rho,\Delta_r,\Delta_\phi)$ can be
projected on these eigenvalues using the following projection
coefficients, found by solving the system
$\mathcal{C}\vec{a}=\vec{\Delta}$, where $\mathcal{C}$ is the matrix
with the eigenvectors:
\begin{eqnarray}
a_1 &=& \frac{1}{2 c_s} ( (c_s+v_r) \Delta_\rho - \Delta_r ) \nonumber\\ 
a_2 &=& \frac{1}{2 c_s} ( (c_s-v_r) \Delta_\rho + \Delta_r ) \\ 
a_3 &=& \Delta_\phi - v_\phi \Delta_\rho~~~~~~~~~~~~~~~ \nonumber
\end{eqnarray}

\subsection{Azimuthal direction}
The Jacobian matrix $\mathcal{A}$ is given by:
\begin{equation}
\mathcal{A}=\left( \begin{array}{ccc} 0 & 1 & 0 \\ - v_r v_\phi &
v_\phi & v_r \\ \frac{c_s^2}{r^2}-v_\phi^2 & 0 & 2 v_\phi
\end{array} \right)
\end{equation}
The eigenvalues of this matrix are:
\begin{eqnarray}
\lambda_1 &=& v_\phi - c_s/r \nonumber\\ 
\lambda_2 &=& v_\phi + c_s/r \\
\lambda_3 &=& v_\phi~~~~~~~~~ \nonumber
\end{eqnarray}
The corresponding eigenvectors:
\begin{eqnarray}
\vec{e}_1 &=& \left(1,v_r,v_\phi-c_s/r \right) \nonumber \\ 
\vec{e}_2 &=& \left(1,v_r,v_\phi+c_s/r \right) \\ 
\vec{e}_3 &=& \left(0,1,0\right)~~~~~~~~~~~~ \nonumber
\end{eqnarray}
A vector $\vec{\Delta}\equiv(\Delta_\rho,\Delta_r,\Delta_\phi)$ can be
projected on these eigenvalues using the following projection
coefficients, found by solving the system
$\mathcal{C}\vec{a}=\vec{\Delta}$, where $\mathcal{C}$ is the matrix
with the eigenvectors:
\begin{eqnarray}
a_1 &=& \frac{1}{2 c_s} ( (c_s+r v_\phi) \Delta_\rho - r\Delta_\phi )\nonumber \\ 
a_2 &=& \frac{1}{2 c_s} ( (c_s-r v_\phi) \Delta_\rho + r\Delta_\phi ) \\ 
a_3 &=& \Delta_r - v_r \Delta_\rho~~~~~~~~~~~~~~~~~~~
\nonumber
\end{eqnarray}

\bibliographystyle{aa} 
\bibliography{planet1.bib}

\end{document}